\documentstyle[11pt,aaspp4]{article} 

\def\ea{{\it et al.}~} 
 
\def\eg{{\it e.g.,~}}
\def\ie{{\it i.e.,~}}

\def \msol {\rm{M}$_\odot$} 
\def \mdot {\rm{M}$_\odot$~yr$^{-1}$} 
\def\kms{km~$\rm{s}^{-1}$}

\begin{document}

\title{Bipolar Outflows and the Evolution of Stars}

\author{Adam Frank}

\begin{abstract}

Hypersonic bipolar outflows
are a ubiquitous phenomena associated with both young and highly
evolved stars. Observations of Planetary Nebulae, the nebulae surrounding
Luminous Blue Variables such as $\eta$ Carinae, Wolf Rayet bubbles, the
circumstellar environment of SN 1987A and Young Stellar Objects
all revealed high velocity outflows with a wide range of
shapes. In this paper I review the current state of our theoretical
understanding of these outflows.

Beginning with Planetary Nebulae considerable progress has been made in
understanding bipolar outflows as the result of stellar winds interacting
with the circumstellar environment.
In what has been called the "Generalized Wind Blown
Bubble" (GWBB) scenario, a fast tenuous wind from the central star
expands into a ambient medium with an aspherical (toroidal) density
distribution.  Inertial gradients due to the gaseous 
torus quickly lead to an
expanding prolate or bipolar shell of swept-up gas bounded by strong
shock waves.  Numerical simulations of the GWBB scenario show a
surprisingly rich variety of gasdynamical behavior,
allowing models to recover many of the observed properties of stellar
bipolar outflows including the development of collimated supersonic
jets.

In this paper we review the physics behind the GWBB scenario in detail
and consider its strengths and weakness.  Alternative models involving
MHD processes are also examined.  Applications of these models to each
of the principle classes of stellar bipolar outflow (YSO, PNe, LBV,
SN87A) are then reviewed.  Outstanding issues in the study of bipolar
outflows are considered as are those
questions which arise when the outflows are viewed as a
single class of phenomena occuring across the HR diagram.

\end{abstract}

\clearpage

\section{Introduction} Bipolar outflows and highly collimated jets are
nearly ubiquitous features associated with stellar mass loss.  From
Young Stellar Objects (YSOs) to Luminous Blue Variables (LBVs) and
Planetary Nebulae (PNe) - the stellar cradle to the grave - there exists
clear evidence for collimated gaseous flows in the form of narrow high
velocity streams or extended bipolar lobes (Fig 1.1).  In YSOs, LBVs and
PNe these collimated hypersonic outflows are observed to transport
prodigious amounts of energy and momentum from their central stars -
enough to constitute a significant fraction of the budgets for the
entire system. Thus outflows and jets are likely to play a significant
role in the evolution of their parent stars. It is remarkable that such
different objects, separated by billions of years of evolution and
decades of solar mass, should drive phenomena so similar.  The
similarity of jets and bipolar outflows across the H-R diagram must tell
us something fundamental and quite general about the nature of stellar
evolution as well as the interaction of stars with their environments.

The purpose of this paper is to review our current understanding of
bipolar outflows.  As we will see there is a unique synergy
between theory and observations in this field providing a
window into fundamental processes such as shocks, instabilities,
ionization dynamics and chemistry. These systems are, however, more than
astrophysical laboratories.  The tension between an extensive
multi-wavelength database and increasingly sophisticated theoretical
tools allows bipolar outflows to act as an Archimedean lever yielding
insights directly into the birth and death of stars. 

Bipolar outflows consitute an exciting field of study with rapidly
expanding frontiers.  The goal of this review is to act as a
tutorial in bipolar outflow studies as well as focusing attention
on particularly pressing unanswered questions.

\subsection{Observational Background} Remarkable progress in
understanding bipolar outflows has been achieved in the last two
decades.  Observationally, the triad of morphologic, kinematic and
spectroscopic studies have provided detailed portraits of bipolar
outflows both individually and as a class of astrophysical object.

Morphological studies using both the HST and ground based instruments
reveal bipolar outflows assuming a wide variety of large scale ($L
\approx 10^{17} ~cm$) {\it global} configurations.  Spherical and
elliptical outflows are observed in evolved systems such as PNe,
Wolf-Rayet nebulae and LBVs (Schwarz \ea 1992; Marstson \ea 1994; Nota \ea
1995).  True bipolar outflows, which appear as two opposing lobes joined
at narrow waist (centered on the star) occur in both young and evolved
systems.  The bipolar lobes exhibit different degrees of collimation
ranging from wide figure {\bf 8} shapes to long, narrow jets. Fig 1.2 shows
a sample of PNe and underscores the extraordinary diversity of 
outflow shapes. The appearence of {\it point symmetry}, where all features are
reflected across a central point as in an {\bf S} 
(NGC 5307, Fig 1.2 lower right), is
a particularly intriguing feature to emerge from recent observations.
Morphological studies also reveal an array of small scale features ($L
\le 10^{16} ~cm$) which go by a variety of suggestive labels: ansae,
knots, arcs, and cometary globules (Balick \ea 1994, O'Dell \& Handron
1996).

Kinematical studies of bipolar outflows (using long slit echelle or
Fabret-Perot observations) reveal flow patterns that simultaneously
possess high degrees of symmetry and complexity (Bryce \ea 1997).  The
most important point for this review is the clear presence of {\it
globally axisymmetric expansion patterns} (Corradi \& Schwartz 1992b,
Lada \& Fitch 1996).  The velocities of the bipolar lobes are larger
than local sound speeds indicating that outflow physics
will be dominated by processes inherent to hypersonic shock
waves.

Spectroscopic studies of line emission provide detailed snapshots of
the microphysical state in the outflows.  Ionization of the lobes
can occur in different ways. Low mass young stars are too cool to
produce significant ionizing flux so bipolar outflows associated with
these stars can only be collisionally ionized by shocks. Outflows from
evolved stars and those associated with high mass young stars are,
however, often photoionized by strong stellar UV fluxes.  In addition
to ions, most classes of bipolar outflow show some evidence for the
presence of molecules and molecular chemistry (Latter \ea 1995,
Bachiller \& Gutie\'rrez 1997).  When spectroscopic studies of ionic
and molecular transitions are coupled with shock
emission/photoionization models, physical conditions in the outflows
(density, temperature, ionization and molecular fractions) can be
determined (Hartigan, Morse \& Raymond 1993, Balick \ea 1994, Dopita 1997).

\subsection{Theoretical Background}

Rapid progress has also been made in the theory of bipolar outflows. 
Analytical and numerical studies have recovered many of the observed
features of outflows through what we will call the Generalized Wind
Blown Bubble (GWBB) paradigm.  In this scenario a fast wind from the
central source expands into a strongly aspherical (toroidal)
environment. The interaction of the wind and environment produces an
expanding bubble bounded by strong shocks. The bubble's velocity is highest
in the direction of lowest density. Thus it is the density gradient in
the environment which establishes a preferred axis for the bipolar
lobes. We will explore this mechanism in considerable detail in section
2.2 and 2.3.  Here we simply emphasize that this theoretical picture
has been remarkably successful at explaining the properties of bipolar
outflows in evolved stars.  It has also found some success in explaining
the properties of bipolar outflows in YSOs but there the issues are
more complex.  One aspect of YSOs which muddies the waters is the
presence of both highly collimated jets and wider bipolar molecular
outflows. The GWBB model provides a robust means of collimating jets
(discussed in section 2.4 and 2.5) and may,
therefore, be relevant to both the manifestations of outflows in YSOs.

The GWBB paradigm has been applied to almost all forms of bipolar
outflows. Models which include both hydrodynamics and microphysics have
been able to recover the global morphology, kinematics and 
ionization patterns in many PNe (Frank \& Mellema 1994b) and SN87A
(Martin \& Arnett 1995). Strong correspondences also exist between GWBB
models and the shapes and kinematics of WR nebulae (Garcia-Seguria \& MacLow
1993), LBVs like $\eta$ Carinia (Frank, Balick \& Davidson 1995) and
symbiotic stars like R Aquarii (Henny \& Dyson 1992). In section 3 we
will discuss the applications of the GWBB theory in more detail
including its relevance to YSOs.

The GWBB scenario has enjoyed enough success in describing outflows from
evolved stars that, in this regime at least, one can speak of a {\it
classical} GWBB model. Numerous variations on its basic theme have been
already explored. In an observational science familiarity can breed
contempt and the limits of the GWBB scenario are already being revealed.
 New observational aspects of evolved star bipolar outflows such as
point-symmetry have been discovered which the GWBB scenario may have
difficulty recovering.  Models invoking MHD processes have been proposed
which may be capable of addressing these issues better than the GWBB
model (Chevalier \& Luo 1994). We will discuss these models in section
2.8.  MHD processes are also believed to produce YSO jets via
 accretion disks. These models
may be relevant for PNe (Soker \& Livio 1994) but may not be as useful
for LBVs, WR bubbles and SN87A-type outflows (cf Washimi \ea 1996).

\subsection{Teleology} While bipolar outflows offer an unparalleled view
into many important physical processes the ultimate purpose of their
study is to understand stellar evolution.  The details of stellar birth
and death are often hidden from view, shrouded behind dense dusty
circumstellar veils.  Bipolar outflows and jets are, on the other hand,
easily observed and have dynamical ages long enough to span important
evolutionary transitions in the life of a star. Thus, bipolar outflows
contain a ``fossil'' record of the central star's history.
  The simultaneous development of high resolution
observational tools and detailed numerical models has allowed
unprecedented contact between theory and reality in bipolar outflow
studies.  Thus there exists the possibility of reading the history of an
individual star or class of stars from the outflows.  This is the most exciting
aspect of bipolar outflow studies however currently
it remains in its infancy.  In section 4
we will consider what is required to build accurate stellar wind
paleontological investigations.

A note on definitions is warranted before we venture any further.  In
what follows we explore bipolar outflows primarily as a {\it
hydrodynamic} phenomena.  While we focus on
jets and jet collimation we approach this issue in the context
of the collimation of bipolar outflows as a whole.  I am, therefore,
interested in systems in which both jets and bipolar outflows arise and
examine the question: can both these outflow phenomena be driven by
the same underlying hydrodynamic collimation process.  There are a
number of excellent reviews of MHD collimation of jets
(Pudritz 1991, Shu 1997). Here, however, the focus is first on bipolar outflows
and second on jets.  Thus I will not address the relativistic beams
associated with neutron stars and black holes. It is quite
possible that the hydrodynamic mechanisms discussed in this review will
have some relevance to those systems as well (Eulderink \& Mellema 1994).

\section{Theory} In this section we review the basic theory of bipolar
wind blown bubbles (WBBs). We begin with a star embedded in some form of
unmagnetized ambient medium.  The star produces a strong wind which
expands into the ambient gas.  To follow the interaction of the wind and
ambient material we must model a multi-dimensional, time-dependent
gaseous flow. We should include heating and cooling of the gas due to
radiation.  A proper treatment of energy source terms requires
calculation of the microphysical state (ionization, chemistry, level
populations). Thus a complete solution of the bipolar WBB problem
requires solving the radiation-gasdynamic equations, \ie the Euler
equations, coupled to radiation transfer and microphysical rate
equations. In what follows we will not write 
expressions for the transfer of radiation or the rate equations for
chemistry, ionization and level populations. The form of these equations
depends too strongly on assumptions made in each environment. Instead we
have used an asterisk to denote those terms which depend on the
radiation field and/or microphysical state (Schmidt-Voight \& Koeppen 1984,
Frank \& Mellma 1994a, Suttner \ea 1997).

The Euler equations take the form (Shu 1994)

\begin{equation} {\partial\rho \over \partial t} + {\nabla} \cdot \rho 
{\bf u} = 0\,, 
\label{masscon} 
\end{equation} 
\begin{equation}
{\partial\rho  {\bf u} \over \partial t} + { \nabla} \cdot \rho  {\bf
uu} = -\nabla P^* \,,
 \label{momcon} 
\end{equation} 
\begin{equation}
{\partial E \over \partial t} +  { \nabla} \cdot  { u} (E + P^*) = H^* -
C^*\,, 
\label{encon} 
\end{equation} 

where 

\begin{equation} E = {1 \over
2} \rho \vert  {\bf u} \vert^2 + {P^* \over (\gamma^* - 1)}\,,
\label{endef} 
\end{equation} 

\noindent and 

\begin{equation} P = {\rho k
T\over \bar \mu^*} \,. 
\label{pdef} 
\end{equation}

\noindent In the above equations $\rho, {\bf v}, P, T$ and $E$ are the
density, velocity, pressure, temperature and energy density
respectively. $\bar{\mu}$ is the mean mass per particle and $\gamma$ is the ratio
of specfic heats. $H$ and $C$ are the
volumetric heating and cooling rates.

We have also ignored processes involving heat conduction, radiation
pressure on grains and molecular and turbulent viscosity.  These process
may or may not be important depending on the environment (Soker 1994).

\subsection{Spherical Bubbles: Analytical Models}

We first consider spherical wind blown bubbles to clarify basic
dynamical issues.  When both the wind and the ambient medium are
isotropic the Euler equations reduce to a system of PDEs in time and
one spacial dimensional (spherical radius $R$). In what follows I borrow
liberally from a number of excellent treatments of spherical WBB theory
(Pikel'ner 1968, Dyson \& de Vries 1972, 
Weaver \ea 1977, Kwok \ea 1978, Dyson \& Williams 1981, Kahn 1983, Koo
\& McKee 1992)

When a stellar wind ``turns on'' it initially expands ballistically
until enough ambient material is swept up for significant momentum to be
exchanged between the two fluids.  A triplet of hydrodynamic
discontinuities then forms defining an ``interaction region'' bounded
internally (externally) by undisturbed wind (ambient) gas.  One can
imagine the interaction region as a spherical shock wave layer cake. At
the outer boundary is an outward facing shock. It accelerates,
compresses and heats ambient material as it propagates. We refer to this
feature as the {\it ambient shock} and denote its position as $R_{as}$.
The inner boundary is defined by an inward facing shock which
decelerates, compresses and heats the stellar wind.  We refer to this
feature as the {\it wind shock}.  Its position is $R_{ws}$.  A {\it contact
discontinuity} (CD), $R_{cd}$, separates the shocked wind and shocked
ambient material.  In the 1 dimensional (1-D) bubble these 
discontinuities form a sequence in radius: $R_{ws} < R_{cd} < R_{as}$ (Fig 2.1).

The compressed gas behind either or both shocks will emit strongly in
optical, UV and IR wavelengths producing a bright shell which defines
the observable "bubble".  The dynamics of the bubble and its emission
characteristics are defined by the strength of post-shock shock cooling.
 Here we consider radiative processes due to collisional excitation to
be the dominant cooling processes. Behind each shock we can define a
cooling timescale $t_c = E_t/\dot{E_t}$, where $E_t$ is the thermal
energy density of the gas.  To first order, radiative heating, often from
a stellar radiation field, contributes only by establishing a ``floor''
in the temperature. We therefore ignore heating in what follows. Radiative cooling can
be expressed in terms of a cooling curve: 
$\dot{E}_t = C(T) = n^2\Lambda(T)$ where
$n$ is the number density of the gas and function $\Lambda(T)$ is a 
sum over many radiative processes emitting a variety of wavelengths. 
The bubble has both cooling and dynamical timescales defined as

\begin{eqnarray} t_c = {3  k T \over 2 n \Lambda(T)}\\ 
t_d = {R_{as}
\over V_{as}} 
\label{bas}
\end{eqnarray}

\noindent where $V_{as}$ is the speed of the ambient shock and we have
assumed (as we shall throughout this paper) that $\gamma = 5/3$.  Comparison
of $t_c$ and $t_d$ separates WBBs into two classes:
Radiative (also known as momentum conserving) and Adiabatic (a.k.a.
energy conserving).

Understanding bubble dynamics requires considering each shock
separately. If $t_c > t_d$ then the gas behind a shock does not have
time to cool before the bubble evolves appreciatively.  If cooling is
weak the gas retains thermal energy gained after passing through the
shock transition and we refer to shock as {\it adiabatic}.  Shu (1994)
has correctly pointed out that a better term would be {\it
non-radiative} since energy not entropy is the issue. We
 will bend to convention, however, and continue using the label
 Adiabatic.  High pressure behind an adiabatic shock limits the post
shock gas compression ($\rho_{post} \le 4 \rho_{pre})$.  In some cases
this can lead to a large separation between the shock and the CD.

If $t_c < t_d$ then the gas cools quickly relative to the bubble's
growth. In this case the shock is {\it radiative}.  The loss of thermal
energy behind a radiative shock means the loss of pressure support as well. 
The shock collapses back towards the contact discontinuity (in a frame
moving with the shock) producing a
thin dense shell. There is no intrinsic limit to the post shock
compression behind a non-magnetized radiative shock (Hollenbach \& McKee 1989).

In almost all cases of interest in this review the densities in the ambient
medium are high enough to ensure that a WBB's ambient shock will be
radiative.  Thus $R_{as} \approx R_{cd}$ and the bubble will have a
thin outer shell.  The properties of the wind, and the wind shock, can however
vary appreciatively from one application to another. The nature of wind
shock will depend upon the wind's speed $V_w$ and mass loss  rate
$\dot{M}_w$.  The density in a steady wind takes the form 

\begin{equation}
\rho_w(R) = {\dot{M}_w \over 4\pi R^2 V_w}.
\label{r2law} 
\end{equation}

Thus winds with high $V_w$ and low $\dot{M}$
are tenuous.  Since $\rho_{pre} = \rho_w(R_{ws})$ the cooling timescale
for shocked wind material can be quite long.  In addition, winds with
high $V_w$ produce high post-shock temperatures. If the wind shock
expands slowly then $V_{ws} \approx V_w$ in the wind's reference frame. 
The post-shock temperature is then

\begin{equation} 
T_{post} = {3 \mu \over 16 k} V_{w}^2
\label{tshock}  
\end{equation}

\noindent which further extends the cooling time.  When $t_c >> t_d$
for material behind the wind shock, the bubble interior fills with hot gas.
Thus an adiabatic wind shock produces an {\it energy conserving bubble}.
 We note that in the literature the terms energy conserving bubble and
adiabatic bubble are often used interchangeably. In the co-moving frame
of the CD the wind shock is
pushed back towards the star, $R_{ws} << R_{cd}$.  In energy conserving
bubbles it is the thermal energy (pressure) of the shocked wind 
which drives the expansion of the bubble as a whole.  The region between
$R_{ws}$ and $R_{cd}$ is often refered to as ``the hot bubble''.

In the other extreme the shocked gas can cool effectively and the wind
shock collapses onto the contact discontinuity, $R_{ws} \approx R_{cd}$.
In this case the bubble consists a double layered thin shell. The
expansion is now driven directly by the ram pressure of
the stellar wind $\rho_w V_w^2$ and the bubble is referred to as momentum
conserving or radiative.

The expansion speed of energy conserving (adiabatic) and momentum
conserving (radiative) bubbles can be easily determined.  We begin with
a spherically symmetric environment characterized as

\begin{equation} 
\rho(t) = \rho_{01} R^{-l} 
\label{amden} 
\end{equation}

\noindent where $\rho_{01}$ is a constant determined by the parameters
of the environment.  In a uniform environment ($l=0$) $\rho_{01}$ is the
actual density (Koo \& McKee 1992).  For an environment created by a
previously deposited wind, ($l=2$) $\rho_{01}$ is proportional to the 
mass loss rate
in the wind divided by the wind speed (equation \ref{r2law}).
 From dimensional arguments alone
(Shore 1995) it is easy to show that an energy conserving bubble driven
by a stellar wind with mechanical luminosity $L_w = .5 \dot{M}_w V_w^2$
expands as

\begin{equation} 
R_{as}^{E}(t) \propto ({L_w \over \rho_{01}})^{({1
\over 5 - l})} t^{({3 \over 5 - l})} 
\end{equation}

One can also show that a momentum conserving bubble driven by a stellar
wind with momentum input $\dot{\Pi} = \dot{M}_w V_w$ expands as

\begin{equation} 
R_{as}^{M}(t) \propto ({\dot{\Pi} \over
\rho_{01}})^{({1 \over 4 - l})} t^{({2 \over 4 - l})} 
\end{equation}

\noindent Notice that $R_{as}$ increases more rapidly for the energy
conserving bubble. Thus the energy lost to radiation in the momentum
conserving case strongly effects bubble dynamics. We will see this again
when we consider bipolar bubbles.

When the bubble is radiative, the wind and ambient shocks lie close to
each other and they expand with the same speed.  In the adiabatic case
the wind shock expands more slowly.  The wind shock's location is
determined by the balance of wind ram pressure and hot bubble thermal
pressure.  Dimensional arguments can again be employed to find the
expansion speed of the wind shock

\begin{equation} R_{ws}^{E}(t) \propto ({L_w \over \rho_{01}})^{({3
\over 10 - 2l})} t^{({4 - l \over 10 - 2l})}. \end{equation}

Koo \& McKee 1992 have shown that WBBs can evolve through a sequence of
configurations from fully radiative to fully adiabatic.  Most bubbles
of interest in this review begin in the radiative regime.  As the
bubble expands the density of the wind at $R_{ws}$ becomes low enough
to push $t_c > t_d$.  Instead of making a direct transition to an
energy conserving bubble, Koo \& McKee 1992 conjectured that another
evolutionary state lay between the adiabatic and radiative
configuration which they called the Partially Radiative Bubble (PRB).
In a PRB the cooling time for the gas is shorter than the age of the
bubble but longer than the time it takes for the unshocked wind to
reach the wind shock i.e. $t_{\rm c} < t_{\rm cross} < t_{\rm d}$ where
$t_{\rm cross} = R_{\rm as}/V_{\rm w}$.  In the PRB stage most of the
shocked wind will have cooled but material which has recently passed
through the shock will remain hot enough to keep $R_{\rm cd} >> R_{\rm
ws}$ (Fig 2.1). While the existence of PRB's has yet to be confirmed in
numerical simulations, Koo \& McKee's paper is highly recommended as it
provides a clear analytic map of the cooling and dynamical evolution in
spherical WBBs.

\subsection{Bipolar Bubbles: Analytical Models}

The WBB paradigm can be extended to embrace elliptical and bipolar
nebulae by generalizing the model to include an aspherical
environmental density distribution $\rho = \rho(R,\theta)$. We imagine
that $\rho(R,\theta)$ describes a toroidal density distribution with a
equator to pole density contrast defined as $q =
\rho(0^o)/ \rho(90^o) = \rho_e / \rho_p$. We always take $q \ge 1$.  In
what follows we consider axisymmetric bubbles ($2-{1 \over 2}$
Dimensions or 2.5-D). When an isotropic stellar wind encounters the
gaseous toroid {\it inertial gradients} (as opposed to pressure) allows
the ambient shock to expand more rapidly along the poles (Fig 2.2). To
see the dominant role of inertia consider an isothermal toroidal ambient
medium.  The highest pressures in environment are achieved in the
equator where $P_{\rm a} \propto \rho_{\rm a} T_{\rm a}$. The
temperature in the environment will, in general, be low ($T_a < 10^4
~K$).  Thus the pressure at the equator will always be orders of
magnitude lower than the driving ram pressure or thermal pressure
achieved in the hot bubble with $P_{\rm hb} \propto \rho_{\rm w}
{V_{\rm w}}^2$.  Only the inertia of the ambient medium will
affect the shape of the bubble.  In his study of energy conserving GWBB
dynamics (in the context of interacting stellar winds) Icke (1988) 
derived an expression for the evolution
of the ambient shock geometry $R_{as} = R_{as}(\theta,t)$.

\begin{equation} 
{\partial R_{\rm as} \over \partial t} = {\lbrace A
(1+({1 \over R_{\rm as}}{\partial R_{\rm as} \over \partial \theta}
)^2)\rbrace}^{1 \over 2} 
\label{Komp} 
\end{equation}

\noindent where the above expression comes from Kompaneets' (1960) formalism and 

\begin{equation} 
A = {\gamma + 1 \over 2} {P_{\rm hb} \over \rho_{01}
(\theta)} 
\label{accpar}
\end{equation}

\noindent $\rho_{01}(\theta)$ describes the angular variation of the
ambient density.  Equation \ref{Komp} shows that $A = A(\theta)$ can be
defined as a local acceleration parameter for the ambient shock.
Therefore the run of $A(\theta)$ determines the asphericity of the
bubble. The Kompaneets approximation assumes that the hot bubble is {\it
isobaric}, i.e. $P_{\rm hb}$ is constant in the region between $R_{ws}$
and $R_{cd}$.  Thus, the angular variation of density (inertia)
in the ambient medium,
$\rho_{01}(\theta)$, determines the angular dependence of $A$ and,
therefore, the shape of the bubble.

In general, analytical determination of the properties of aspherical
WBBs is difficult.  The problem involves partial differential equations
which must be solved in two spatial dimensions and time.  Dyson (1977)
and Kahn \& West (1985) provided the analytical solutions for bipolar energy
conserving nebula using the GWBB paradigm. Icke's 1988 investigation
explored a wider range of solutions demonstrating that strongly
collimated solutions for $R_{as}(\theta,t)$ could be generated  if the
torus had a narrow opening angle.  In a study of R Aquarii, Henny \&
Dyson (1992) used a model based on momentum conservation in the shell
and included emission characteristics of the bubble for both shock excited and
photo-ionized emission.  Working off a novel method proposed by
Giuliani (1982), Dwarkadas, Chevalier and Blondin (1996) produced similarity
solutions for bipolar WBBs. These solutions were quite powerful in that
they predicted not only the shape of the bubble, but also the mass
motions along the shell of swept up ambient gas.  
Tangential motions along the shell can lead to substantial modifications 
of shell density (Kahn \& West 1985, Wang \& Mazzali 1992)

Any study of the GWBB formalism must include specification of the
ambient density distribution.  The physics which generates asphericity
in the environment will depend on the application and we will discuss
specific mechanisms in more detail in section 3.6.  Here we note that
most investigations have used ad-hoc seperable functions to control the
ambient density distribution $\rho(r,\theta) \propto R^{-l}~F(\theta)$.
Three examples which have been used extensively in the literature are
the Icke, Inverse Icke and Luo \& McCray functions (Icke, Balick \&
Preston 1989, Luo \& McCray 1991, Dwarkadas, Chevalier and Blondin 1996). 
These take the form

\begin{equation} 
F_{icke}(\theta) = [ 1 - \alpha + \alpha
e^{(\beta\cos^2\theta- \beta)} ]^{-1} 
\label{ad1} 
\end{equation}
\begin{equation} 
F_{inv icke}(\theta) = C[ 1 - \alpha{e^{(-
2\beta\cos^2\theta)-1)} \over e^{(-2\beta-1)}}] 
\label{ad2}
\end{equation} 
\begin{equation} 
F_{LM}(\theta) = {3 \over 3-\alpha} (1 -
\alpha \cos^2\theta). 
\label{ad3} 
\end{equation}

\noindent The parameter $\alpha$ controls the pole to equator density
contrast (which in turn defines $q$) while $\beta$ controls the
steepness of the turnover from $\rho_p$ to $\rho_e$.  There is no
physics inherent to eqs \ref{ad1} - \ref{ad3}.  They are simply useful for
controlling the shape of the ambient medium.  More recent studies
(Garcia-Segura \ea 1997, Collins \ea 1998) have attempted to
incorporate specific physical models for the creation of the toroidal
environment into GWBB calculations.  This is one of the most important paths
for future research efforts to follow.

\subsection{Numerical Models}

In this section we will consider multi-dimensional numerical models of
WBBs. The inherent complexity of the GWBB problem make simulations a
necessary tool for explicating the true range of hydrodynamic flow
patterns possible in these systems. The evolution of 2.5-D bipolar
outflows under the GWBB scenario has now been well studied with computer
models.   We note that 1-D spherically symmetric numerical models that
have also been extensively explored.  These models have reached a fairly
high degree of sophistication in terms of their treatments of both
hydrodynamics and microphysical processes.  The interested reader should
consult the following references (Schmidt-Voight \& Koeppen, 1987; Marten \&
Schoneberner 1991; Frank 1994; Mellema 1994; Arthur, Henny \& Dyson 1996)

The first and most frequent application of numerical GWBB
models has been to PNe. In this section we focus on these studies since
PNe constitute the archetypical interacting stellar wind bipolar bubble.
In PNe the ambient density distribution is assumed to form from a slow,
($V_a \approx 10$ \kms), dense, ($\dot{M}_a \approx 10^{-5}$ \mdot), wind
expelled when the central star was on the Asymptotic Giant Branch (AGB),
$l = 2$ in equation
\ref{amden}). The stellar wind has a high velocity, 
($V_w \approx 10^3$ \kms),
\kms and low mass loss rate, ($\dot{M}_w \approx 10^{-7}$ \mdot). Soker \&
Livio (1989) were the first to explore the time-dependent evolution of
bipolar PNe with the GWBB formalism.  They confirmed the ability of the paradigm to produce
bipolar bubbles in time-dependent calculations.  Using higher order
methods and higher resolution grids Mellema, Eulderink \& Icke (1991) and
Icke, Balick \& Frank (1992) explored the hydrodynamic flow pattern in
bipolar wind blown bubbles in more detail.  Their results provided a
more extensive mapping of parameter space articulating the dynamics of
the ambient/wind shocks in greater detail.  The formation of highly
collimated jets in the hot bubble was one of the most surprising results
of these simulations.  While it is now recognized that PNe are also 
jet-bearing systems (Livio \& Soker 1994) these simulations 
showed how jets could effectively form within PNe WBBs (Rozyczka \& 
Tenorio-Tagle 1985). We will address the issue of jet collimation again 
in the following
sections.

Radiative losses were not included in the simulations cited above.  Thus
both the wind and ambient shocks were adiabatic.  Given the
high wind speeds this was 
appropriate only for the wind shock . In
addition, without radiative losses these simulations could not be
directly compared with observations.  In a series of papers by Frank \&
Mellema (Frank \& Mellema 1994a, Frank \& Mellema 1994b, Mellema \&
Frank 1995, Mellema 1995) a numerical code was developed which included
hydrodynamics, microphysics and radiation transfer from the hot central
star.  These models tracked cooling and emission behind the
ambient shock.  Fig 2.3 shows a series of simulations from Frank \&
Mellema (1994) which illustrates the dependence of outflow geometry on
initial conditions.  Fig 2.3 displays the evolution of density in four
PNe simulations.  The equator to pole density contrast, $q$, increases
as one moves down the figure.  For low values of $q$, ($q < 2$), the
bubble becomes mildly elliptical. Intermediate values of $q$, ($2 < q <
5$), produce distinct equatorial and polar regions of the ambient shock;
\ie bipolar lobes develop.  Larger values of $q$ produce bubbles that
are highly collimated.  Fig 2.3 demonstrates the ability of the
 GWBB model to produce
different bubble shapes. The proof of the pudding is in the
confrontation with observations and we will consider that aspect of the
problem in the section on PNe (3.1).  For now we simply state that these
models recover many of the observed morphological, kinematic and
ionization structures seen in real bipolar WBBs.

The expansion speed of AGB slow winds is often of the same magnitude as
the ambient shock speed at the equator ($20 - 40$ \kms). This is true of
most situations where a WBB forms from a cooler star evolving into hot
star (\eg SN87A).   Dwarkadas, Chevalier \& Blondin (1996) recognized that 
this presented a problem and explored the effect of slow wind speeds on bipolar WBB
evolution.  Their results demonstrated that the effects of a density
contrast in the ambient medium are washed out if the ambient gas is
expanding too fast.  When the ambient shock and slow wind velocities
are comparable, bubbles tend towards spherical or elliptical
configurations.  This occurs because the bubble shell is always
``catching up'' to the slow wind and no inertial constraint can be
imposed.

Dwarkadas. Chevalier \& Blondin also provided the most detailed
comparison between numerical simulations and analytical models
available so far.  Their results confirm that the two approaches yield
similar bubble shapes as long as the gas in the ``hot bubble'' remaines
isobaric (Fig 2.4).  Deviations from the isobaric condition produces
bubbles which are narrower and more collimated than self-similar
solutions predict.  In most bipolar WBB models departures from the
isobaric condition could be traced to the development of an aspherical
wind shock. Asphericity in the wind shock was not uniformly anticipated
by researchers developing analytical models. Most analytical studies
have assumed a spherical wind shock while almost all simulations of
bipolar WBBs produce wind shock's with prolate geometries.  Prolate
wind shocks can be seen, for instance, in the all but the first
simulation presented in Fig 2.3. The implications of aspherical wind
shocks for bipolar WBBs dynamics has proven to be profound as we
discuss in the next section.

\subsection{Jet Collimation: Shock Focused Inertial Confinement}

Hydrodynamic jet collimation has, for the most part, fallen out of
favor. Currently the consensus in the astrophysical community holds that
jets are collimated via MHD processes associated with accretion disks
(K\"onigl 1989, Pudritz 1991, Shu 1997).  Previous attempts to develop
hydrodynamic jet collimation mechanisms relied primarily on deLaval
nozzles (K\"onigl 1982, Smith \ea 1983).   A deLaval nozzle is a means of
turning random thermal motions into bulk supersonic flow by forcing hot
gas to pass through a narrow constriction (Shu 1994).

The simulations discussed in the last section revived the issue of
hydrodynamic jet collimation by demonstrating that jets could be
effectively generated in bipolar WBBs.  The collimation relies on the
development of an aspherical wind shock to focus post-shock streamlines
towards the axis.  The role of the inner shock as a hydrodynamic lens
was anticipated in analytical studies by both Eichler 1982 and Icke 1988.
The process seen in the simulations, termed Shock Focused Inertial
Confinement (SFIC), appears to be one of the most effective and robust
hydrodynamic collimation mechanisms currently known (Icke \ea 1992).  It
has been explored in a number of contexts including low and high mass
YSOs (Section 3.4; Frank \& Mellema 1996, Mellema \& Frank 1998, York 
\& Welz 1996) and relativistic jets from AGN (Eulderink \& Mellema 1994).

The SFIC mechanism relies primarily on the behavior of oblique shock
waves.  When the inward facing wind shock assumes a prolate geometry
the radially streaming wind encounters the shock face at an oblique
angle.  Only the normal component of the wind velocity feels the effect
of the shock.  Tangential components of the velocity remain unchanged.
In this way post-shock velocity vectors are tipped away from the shock
normal and wind material is focused into a collimated beam.  Icke (1988)
and Frank \& Mellema (1996) presented analytical models for this
process under the assumption of an elliptical inner shock.  Their
results demonstrated that even mildly aspherical wind shocks could
produce significant flow focusing.

One of the most striking results of the analytical models was the
prediction that {\it supersonic} post shock speeds can be achieved with
out a delaval nozzle.  Frank \& Mellema dervied equations for the Mach
number behind an prolate elliptical wind shock. Their results showed
that supersonic collimated flows occur behind strongly aspherical
adiabatic wind shocks.  They also found that even mildly aspherical
wind shocks can produce supersonic focusing if the post-shock flows
cool effectively. This means that supersonic highly collimated jets can
be formed even when no deLaval nozzles are present.

In Fig 2.5 we present results of a SFIC jet collimation simulation from
Mellema \& Frank 1997. In these simulations, appropriate to YSOs, a
spherical wind is driven into a stationary gaseous torus. Radiation
losses are included.  The wind shock, apparent at
the base of the flow, is strongly aspherical.  This produces
supersonic post-shock flow and initiates shock
focusing which can be seen in a velocity vectors near the top of the wind
shock.  The overall collimation of the shocked wind into a narrow jet is
clearly evident.  Note the presence of shocks within the beam which
indicates that the jet is supersonic.  The internal shocks are  expected
from the classical theory of supersonic jet dynamics (Norman 1993).  

The bubble shown in Fig 2.5 is effectively energy conserving and the jet
is composed of relatively hot shocked wind gas. Since the jet is
supersonic the role of thermal energy in driving the bubble is relatively
unimportant and the dynamics of the global flow pattern lies between
that of a jet and a WBB.

\subsection{Jet Collimation: Converging Conical Flows}

An additional hydrodynamic collimating mechanism becomes available in
momentum conserving bubbles.  When the wind shock is radiative post-shock 
material is compressed into a thin shell bounded on the outside by
the contact discontinuity.  If the bubble becomes prolate, wind material
which passes through the inner shock is focused towards the poles as in
the SFIC mechanism.  In a prolate radiative bubble, however, the narrow
width of the shell constrains the post-shock flow. Material is forced to
stream along the contact discontinuity forming a {\it converging conical
flow} as it reaches the poles. The opposing streams meet directly over
the poles producing a shock that redirects gas into a jet.  Fig 2.6
presents a schematic of the process. Since we are considering
radiative shocks, converging conical flows will produce
dense cold jets.

This process was first investigated by Cant{\'o} \& Rodr{\'\i}guez (1980).
Smith (1986) studied a similar version of the idea.  Cant{\'o},
Tenorio-Tagle \&\ R{\'o}\.zyczka (1988) and Tenorio-Tagle \ea (1988) used
idealized analytical and numerical models treating the converging beams
as initial conditions to explore the jet formation  in greater detail.
Peter \& Eichler (1996) have performed similar simulations in the context
of converging disk winds.  The simulations of Mellema \& Frank (1997)
placed the development of these flows in the context of global bipolar
WBB evolution.  Their simulations demonstrated that such flows
 are a natural and robust
consequence of the interaction between a wind and the surrounding
material.

Fig 2.7 shows a the results of Borkowski, Blondin \& Harington (1997) who
performed simulations of a wind expanding into a pressure stratified
environment.  Their models were targeted at explaining the jet bearing
PNe He3-1475.  Fig 2.7 shows that jets form as a natural consequence
of aspherical radiative WBBs.  

Recall that as a bubble expands the density in the wind immediately
interior to the shock decreases as $\rho_w(R_{sw}) \propto R_{sw}^{-2}$
(eqn. \ref{r2law}). Thus at some finite radius the wind shock will
change from momentum conserving to energy conserving. If the geometry
of the ambient medium takes the right form one can expect a era of jet
collimation via converging conical flows followed by an energy
conserving phase during which the SFIC mechanism may operate. Such a
transition was observed in the models of Mellema \& Frank (1997) and
has been proposed to explain jets and ansae in PNe by Frank, Balick \&
Livio (1996).  Recent calculations by Dwarkadas \& Balick (1998a) have
shown that for some evolving winds cases instabilities along
the shell can disrupt the converging flow.

While the studies cited above have all demonstrated that converging
conical flows are a promising jet collimation mechanism there is an
important caveat. It is not yet clear if the flow patterns are an
artifact of the 2.5-D geometry used the simulations.  In axisymmetric
models the axis is a rigid boundary. Converging flows have no choice but
to refect off the axis.  In 3-D the convergence point could be unstable.
 If material is not perfectly focused to a point a time-variable spray
rather than a tightly collimated beam may result. It is therefore
important not to be carried away by  highly symmetric results from
highly symmetric models. The final judgement will require 3-D models. 
Care must be taken here as well. Low resolution 3-D models may not
adequately answer the question as material will have to fragment on size
scales larger than a computational cell. Achieving high resolution will
be particularly important for determining the efficacy of radiative
hydrodynamic collimation mechanisms.

\subsection{Development of the Toroidal Environment} The GWBB model
relies on the presence of an aspherical environmental density
distribution - a dense torus.  There is evidence for the existence
of such distribution in both young and evolved stars.  Toroidal
distributions with high pole to equator contrasts have been observed in
the predecessors of PNe (Meixner \ea 1997). It should be noted however, that
many AGB stars show spherical circumstellar envelopes up until the very end
of the AGB phase. The existence of disks and
flattened density distributions surrounding YSOs is also well established
(McCaughrean \& O'Dell 1996, Cabrit \ea 1996).

For YSOs the orgin of toroidal density distributions is easily explained
via the collapse of a rotating or magnetized cloud or the collapse of a
filament (Tereby, Cassin \& Shu 1984, Li \& Shu 1996, Hartmann, Calvet \&
Boss 1996). The density distribution in these systems will take the form
of a thin disk on small scales with thicker toroids perhaps existing 
on larger scales (Close \ea 1997).  For evolved stars however, the 
origin of the toroids poses a serious and interesting challenge.  A number of {\it single star}
processes have been proposed including dust formation instabilities
(Dorfi \& Hoefner 1996) and star spots (Frank 1995, Soker 1998).  These mechanisms
almost all rely on rotation in one way or another (Asida \& Tuchman 1995)

The Wind Compressed Disk (WCD) mechanism of Bjorkman \& Cassinelli 1992 has been
one of the most promising single star models explored to date (Owocki, Cranmer
\& Blondin 1994).  WCD formation is effective
for rotating stars.  Wind streamlines are deflected towards the axis in the
competition of radial radiation pressure forces and the coriolis effect. The WCD
mechanism produces an ``excretion disk'' which flows away from the star. Though
the model was originally intended for B[e] stars it has been applied to other
classes of objects including AGB stars (Ignace, Bjorkman \& Cassinelli 1996).  In
most cases the rotation rate of the star needs to be a signifigant fraction of
the critical rotation rate for signifigant wind compression to occur. Garcia-Segura,
Langer \& MacLow (1997) have noted that the critical rotation rate 
depends on the luminosity of the star via

\begin{equation}
\Omega_c = \sqrt{G M_* (\Gamma - 1) \over R_*^3}
\label{omcrit}
\end{equation}

\noindent where $\Gamma = L_*/L_{Ed}$ and $L_{Ed}$ is the Eddington luminosity.
$\Omega_c$ may approach the actual rotation rate of the star as its
luminosity increases .  Thus the WCD mechanism may be particularly effective for
stars close to their Eddington limit (LBVs). 

The greatest advantage of the WCD model is that it links the properties
of the stellar progenitor directly to the bipolar bubble.  The WCD model
produces density and velocity distributions for the ambient medium
which can be fed directly into GWBB models (Garcia-Sequra, Langer \&
MacLow 1997; Collins \ea 1998).  Recent research has however casts some doubt
on the effectivelness of the WCD mechanism in line driven winds (Owocki
\ea 1996).  Its fate in other contexts remain a important area of
research.

The work of Livio, Soker and collaborators (Livio \& Soker 1988; 
 Soker \& Livio 1994; Rasio \& Livio 1996; Soker 1997) has provided
strong support for the role of binary companions in producing the
equatorial density enhancements. As these authors have shown, red giant
rotation rates are too low to allow any of the mechanisms explored to
date to produce dense slowly expanding toroids. Thus spin-up of the
stars is most likely required to produce the required rotation speeds
for mechanisms such as the WCD model. This can be accomplished via tidal
interactions or the spiraling-in of a companion in a common envelope
phase.  A binary merger can also dump enough energy into the atmosphere
of the primary to eject it.  Numerical models of the process demonstrate
that much of the envelope mass is ejected into the orbital plane of the
binary producing the required pole to equator density contrast (Terman,
Taam \& Hernquist 1994, 1995, Sandquist \ea 1998).  Soker (1997a) has concluded that the
distinction between binary and single star models may be a semantic
since even brown dwarf companions or giant planets can lead to aspherical
environments.  In particular, Soker 1996 has shown that a Jupiter sized
planet could contribute enough spin-up to distort a giant star wind into
a configuration whereby the GWBB scenario would produce an elliptical
nebula.

How important are binary companions for the formation of a slow torus?
This is one of the critical issues facing bipolar ouflow studies.
While there is obervational evidence for the presence of binaries in
bipolar PNe (Pollacco 1998) it is not yet clear if single star bipolars
greatly outnumber binary bipolars.  The lack of sufficent binaries may
be answered by appeals to coalesence in a common envelope phase. If
this is true it would be useful to find an obervational signature of
coalesence in the central star, \ie do PNe central stars which are the
product of a merger look different from the products of single star
evolution? 

Another issue which must be addressed is the speed of the
slow torus in a binary ejection model.  Simulations seem to indicate
that the when the giant's envelope is ejected in a binary merger the
highest speeds are obtained at the equator.  As Dwarkadas, Chevalier \&
Blondin have shown, this could reduce the effectiveness of the GWBB
process. In spite of these problems, it appears that the
presence of some form of companion will be required to produce slow
torii and, hence, bipolar outflows.

\subsection{Magnetohydrodynamic Models} Thus far we have focused on
models where the shaping of the bipolar bubbles is accomplished solely
though hydrodynamic forces. The development
of a preferred axis for the bubble's expansion can, of course, occur in
other ways. The most likely alternative is a magnetic field. Pascoli
(1985, 1997) and Gurzadyan (1996) have both proposed models where fields
embedded in the ambient medium (a previously ejected wind) produce the
bipolar morphologies.  Evidence for the existence of these fields
remains ambiguous. Motivated by the desire to find a model which could
produce bipolar outflows without the need of a companion star Chevalier
\& Luo (1994) developed a model which relied on hydromagnetic forces in
the stellar wind itself.  Using the Sun as a
template they argued that rotation of the star produces a stellar wind
with a dominantly toroidal field (${\bf B} = B_\phi {\bf e}_\phi$). When the
magnetized wind passes through the wind shock compression strengthens
the field.  If the initial field is strong enough, magnetic pinch forces,
also known as {\it hoop stresses}, from the toroidal field
constrain the expansion of the bubble along the equator. No such force
inhibits the bubble along the poles and a bipolar bubble
develops without a dense torus.

A critical parameter for these models is the ratio of magnetic to
kinetic energy in the wind,

\begin{equation} \sigma = {B^2 \over 4\pi \rho_w V_w^2} = {B_*^2 R_*^2
\over \dot{M}_w V_w} ({V_{rot} \over V_w})^2 \label{sigeq}
\end{equation}

\noindent where the $*$ subscripts refer to values at the stellar
surface and $V_{rot}$ is rotation velocity of the star. The expression
on the right hand side of equation \ref{sigeq} is based on the assumed
form of the magnetic field.  Using an analytical formulation of the
problem Chevalier \& Luo (1994) demonstrated that magnetic forces in the
hot bubble (they assumed an adiabatic wind shock) could produce
significant departures from spherical morphologies.  For a strong
bipolar morphology to develop they found that $\sigma > 10^{-4}$ (Soker
1997a).

Axisymmetric (2.5-D) numerical models of this mechanism have been
calculated by Rozyczka \& Franco (1996) and Gracia-Segura \ea (1997).
These models confirmed the feasibility of the Luo \& Chevalier scenario
(Fig 2.8) and demonstrated that realistic bipolar shapes could be
obtained.  New features were also seen including the potential for jet
collimation via the magnetic hoop stresses.   Fully 3-D models have been
presented by Garcia-Sequra 1997.  In these simulations precession of the
star's magnetic axis produced precession in the bubble as a whole
including a jet which formed within the bubble (Fig 2.9). The resulting
shapes were quite similar to the point-symmetric morphologies observed
in some PNe (Schwarz, Corradi \& Melnick 1992).  It is noteworthy however that
Garcia-Sequra was forced to include a slow torus (calculated via the WCD
formalism) in his calculations in order to form narrow jets.  In addition,
a rather particular sequence of two or three winds were invoked to produce 
the desired results.

The {\it Magnetized Wind Bubble} (MWB) models outlined above offer an
strong alternative to the ``classic'' GWBB formalism for producing
bipolar outflows in evolved stars.  If, however, the GWBB scenario has
already proven to be effective one can ask what purpose is served by
invoking magnetic fields.  The question is particularly relevant since
evidence for the existence of strong fields in bipolar WBBs does not yet
exist while the presence of gaseous toroids has been established.  One
of their most attractive features the MWB scenario is its ability to
impose large scale non-axisymetric patterns such as point symmetry on
bipolar bubbles. While it may be possible to achieve such patterns in
purely hydrodynamic flows by invoking a ``wobble'' in the gaseous torus,
magnetic fields are ``stiffer'' and can offer a longer lever arm for
achieving coherence across large length scales.  Other MHD
mechanisms such as collimated accretion disk winds offer similar
advantages ( Soker \& Livio 1994, Livio \& Pringle 1997).

Criticism of the MWB models hinges both on the strength and topology of
the field (Soker 1997a, Livio 1998).  Numerical simulations of the MWB
model have used very high values of the field, $\sigma \approx .01$, to
achieve strong collimation (Garcia-Segura 1997).  As Soker 1997a has
pointed out these values where chosen based on conditions in the sun
where the wind is driven by the magnetic field.  The winds which produce
bipolar WBBs surrounding evolved stars however are radiatively driven
making the sun is a poor template. For bipolar PNe, Chevalier \& Luo
1994 point out that only a small fraction of white-dwarfs show magnetic
fields strong enough to justify the values of $\sigma$ used in the
numerical simulations.

Soker 1997a has also raised the issue of field topology. The MWB model
requires the the field to circle the star in the planes parallel to the
equator.  In the solar wind however this is not the case.  The magnetic
field in the solar wind is composed of distinct sectors across which
the field reverses sign (Livio 1997, Bieber \& Rust 1995, Gosling
1996). A similar problem arises for the direction of the field across
the equatorial plane. Both solar observations and dynamo models show
that $B_\phi$ has opposite polarity between the two hemispheres (Weis
1994).  Thus the field must go through zero at the equator, just where
it is needed most.  In his 3-D models Garcia-Segura (1997) chose
$B_\phi \propto \sin\theta$ giving the field its maximum value at the
equator. Even if the $B_\phi$ transition is narrow the field reversal
at the bubble's equator should be unstable to tearing mode
instabilities and reconnection (Priest 1984).  Diffusive
processes such as reconnection convert magnetic energy needed for
collimation into thermal energy and pressure which would inflate equatorial regions.  

Estimation of the reconnection time in a MWB is difficult because the
processes which drive reconnection are not clearly understood (Lazarian
\& Vishniac 1998).  There is a long standing debate concerning the
existence of ``fast'' reconnection processes (Parker 1979).  In general
the reconnection rate should go as $L/V_r$ where L is a characteristic
size scale and $V_r$ is the speed with which flux can be driven into
the reconnection zone.  When collisional effects alone are considered
the resistivity in cosmic plasmas is very low.  When theoretical models
such as the Sweet-Parker process (Cowley 1985) use these forms for the
resitivity they yield $V_r << V_a$ where $V_a$ is the local alfv\'en
velocity.  While many theoretical arguments produce extremely slow
reconnection speeds observational evidence from solar flares, indicates
that $V_r \sim .1 - .01 V_a$ (Dere 1996, Innes \ea 1997).  If such scalings hold in the hot
bubble of a MWB then the time scale over which magnetic energy in the
equatorial regions will be dissipated, $\tau_e$ can be calculated.
Since the size scale of field reversal region $l_e$, is unknown we will
calculate the timescale $\tau_d$ for all the magentic flux in a MWB of
size $R$ to be convected into and dissipated within the reconnetion
zone,

\begin{equation} \tau_d \approx \epsilon {R \over V_a} \label{td}
\end{equation}

where $\epsilon = 10 ~- ~100$.  Note that unless $l_e << R$ a large
scale gradient in $P_b$ would exist which would negate the fields
ability to produce collimation.  Thus $\tau_d >> \tau_e$ and $\tau_d$
represents a upper limit to the time-scale for reconnection to be
important.

The alfv\'en velocity can be determined from the initial conditions for
the bubble.  Since magnetic forces are expected to be comparable to gas
pressure forces we may assume equipartion such that $\beta = P_g/P_b =
1$.  In this way one may show that

\begin{equation} V_a \approx \sqrt{3 \over 8} V_w \label{va}
\end{equation}

({\it cf} Dagni \& Soker 1998). If we use canonical values of $R \approx 10^{17}$ cm and $V_w = 10^3$
\kms we find $\tau_d \approx 530 - 5300$ y. This is comparible to the
dynamical timescale for many PNe.  Thus it is possible that
reconnection will play an important role in the evolution of a MWB by
changing the field topology and converting magnetic energy into heat
and hence thermal pressure.

\section{Applications}

In the last section we discussed physical processes which shape bipolar
outflows. The physics was generic and could be obtained in any bipolar
WBB.  In what follows we consider applications of these ideas
focusing separately on each class of stellar environment in which bipolar WBB 
occur.

\subsection{Planetary Nebulae: PNe}

PNe have a long history of serving as laboratories for new astrophysical
processes (forbidden lines are the most famous example Osterbrock (1989)).
They have played a similar role in understanding bipolar outflows.  As
we have discussed in preceding sections, PNe are interacting wind systems
- a fast tenuous wind from a hot central star expands into slow dense
wind expelled during the star's AGB phase (Kwok \ea 1979, Kahn 1983).
Kahn \& West 1985 were the first to consider a GWBB model for PNe. 
Compiling a catalog of PNe with various shapes Balick (1987) expanded on
the theme outlining a broad GWBB paradigm which embraced the majority
of PNe observations.  Balick's conjecture was successfully explored for
the first time in Icke, Balick \& Preston (1989) via an analytical Kompaneets 
(1960) formalism. The numerical models discussed section 2.3 confirmed the
validity of the GWBB model for PNe. As was discussed in section 2.6
the equatorial density enhancement in the AGB wind required for Balick's
scenario is likely to occur via the effects of a companion star.

Confrontaion between PNe models and reality became possible with the advent
of radiation-gasdynamic simulations. The simultaneous computation of
hydrodynamics and radiative/microphysical processes allowed {\it
synthetic observations} of the models to be computed by rotating 2.5-D
emissition maps about the symmetry axis and projecting them on the sky
at different inclinations.  Fig 3.1 shows a figure from Frank \ea 1993
where a single PNe model projected at six different angles is compared
with six real PNe.  The rings apparent in the synthetic observations are
high density regions formed where the shock outlining the lobes 
intersects with the shock
defining the equatorial regions of the bubble.  As the projection angle
increases the rings come to dominate the synthetic shapes producing ``eyes'' like
those seen in the Owl nebula. Fig 3.1 demonstrates that when projection
effects are included even a single GWBB model can embrace many of the
observed PNe shapes.

We must note an inherent danger in creating synthetic observations from
2.5-D simulations. When instabilities such as corrugations occur in the
shell of an axisymmetric simulation these features will, upon rotation
and projection, appear as rings.  It must be emphisized that such rings
are entirely unrealistic. There is no apriori reason to believe locally
driven instabilities will take so a coherent a form when allowed 
to evolve in 3 dimensions.  When considering the effect of
instabilities on the {\it appearence} of of bipolar WBBs fully 3-D
models are required.

A successful model must also account for the kinematics of PNe. Fig 3.2
shows a series of synthetic position-velocity (PV) diagrams for the
simulation shown in Fig 3.1. In each of the 6 panels the slit is placed
along the long axis of the bubble.  As the projection angle of the
bubble is increased a skewed figure {\bf 8} pattern emerges.  The skewing of
the PV pattern emerges as  high velocity material at the top (bottom) of
the bubble is blue (red) shifted in projection onto the line of sight. 
Such patterns have been well documented in observations 
(Balick, Preston \& Icke 1987, O'Dell, Weiner \& Chu 1990,
Corradi \& Schwarz 1993abc). Together with synthetic images these
kinematical predictions show that the GWBB model offers quite a good
match to a large subset of the observational database.  One would expect
that alternative models for bipolar PNe offer at least as good
a match without increasing the number of input assumptions.

Schwarz, Corradi \& Melnick (1992) and Manchado \ea (1997) have produced more
extensive catalogs of PNe shapes.  Statistical studies with this
enlarged sample allow inferences to be drawn about the relation between
PNe and their central stars (Stanghellini, Corradi, Schwarz, 1993). One
intriguing conclusion of this work is that strongly bipolar PNe tend to
originate from more massive stars. Such a correlation may originate in
the rapid evolution ($t \approx 10^4$ y) of PNe central stars from a
cool ($T \approx 3\times 10^3$ K) giant to a hot ($T > 3\times 10^4$ K)
dwarf configuration (Schonberner 1986).  Since the ionizing flux
increases with $T$ and the velocity of the central star wind is
radiation driven, the wind evolves
along with the star (Kudritzki, Pauldrach \& Abbott 1989). 
Stellar evolution models show the fastest evolution for
more massive central stars.

Mellema (1995) included the evolution of both the central star and
stellar wind in his numerical models.  During the early stages of
evolution in his simulations, ionization fronts deposited enough energy
and momentum in the AGB wind to change its geometry before the bubble
reached a bipolar configuration.  The ionization of the AGB torus
produces a strong pressure gradient. Gas is driven off the torus
towards the pole decreasing the density contrast (Mellema \& Frank
1995). Mellema (1997) demonstrated that this effect leads to a
morphological segregation between PNe from high and low mass stars.
More massive stars evolve quickly and produce higher velocity winds
relatively early compared with their lower mass cousins.  In the higher
mass case the ambient shock reaches large radii quickly before the
pre-shaping of the environment by ionization is significant.
Thus massive star bubbles experience the full effect of the initial
density contrast in the AGB wind and attain strongly bipolar
configurations.

Rotation of the central star during both its AGB and PNe phase may play
an important role.  In the models of Garcia-Segura \ea (1997) rotation
in the AGB envelope produces a equatorial enhancement via the WCD model
(section 2.6).  Rotation of the star is also necessary to produce the
strong fields needed for the MWB models. The use of the WCD model for
Red Giant stars is a very promising avenue of research which can be
applied to other classes of bipolar WBBs (SN87A, Section 3.3; Collins
\ea 1998).  It is worth noting here however that it not
yet clear how the WCD mechanism will work in giant star envelopes.
Models of AGB winds rely on radiation pressure on dust while the WCD
mechanism was formulated for line-driven winds.  This issue should be
addressed in detail in future research.

Many PNe exhibit well collimated jets or pairs of high velocity knots
aligned along the major axis of the nebula.  Some of these features,
known as Ansae (Aller 1947), are part of a broader class of structures
called FLIERS (Fast moving Low Ionization Emission regions) which exist
in many PNe (Balick \ea 1993, 1994, Hajian \ea 1997). The origin of the jets and ansae is
currently a subject of considerable debate.  Hydrodynamic mechanisms
such as SFIC or converging conical flows have been proposed by Mellema
1996 and Frank, Balick \& Livio (1996).  MHD mechanisms based on the
Chevalier \& Luo MWB model have also been explored (Garcia -Segura
1997).  In Livio and Soker (1994) and Livio \& Pringle (1996) the jets have
been linked to magento-centrifugal processes associated with an
accretion disk.  The disk forms around either the binary companion or
around the primary.  Mellema \ea (1998)  and Hajian (1998) have also
explored a model where ansae result from photoionized clumps of gas
experiencing the Oort-Spitzer rocket effect.

The fact that many jets and ansae appear outside the apparent position
of the ambient shock argues that they formed before the onset of a
classical GWBB phase. The focus then shifts to the elusive
Proto-Planetary Nebulae (PPNe).  These are objects currently making the
transition from AGB to PNe. Many PPNe such as the Egg Nebula (CRL2688
Sahai 1998a, 1998b) already show highly collimated high speed ($V \approx
200$ \kms) outflows.  The occurrence of bipolar outflows so early in the
transition may pose a problem for both the GWBB and MWB models because
the star is too cool to produce a high speed wind.   If magnetized
accretion disks form in these systems it might solve the problem via
some kind of disk wind. There are however examples of PNe which show
jets forming from an evacuated cavity where no disk is apparent (He3-1357;
Bobrowsky \ea 1998). It is also noteworthy that much of the AGB mass
loss surrounding bipolar PPNe appears spherically symmetric.  This
indicates that the toroidal wind or accretion disk occurs very late in
the AGB phase  (Livio 1997).  The origin of bipolar PPNe outflows
constitutes one of the most important issues currently facing PNe and
bipolar outflow studies.

In the years since Balick published his catalog it has become clear that
point-symmetry is a characteristic in many PNe.   The {\bf S} shaped
morphology of the jet bearing PNe Fl 1 (Lopez, Meaburn, \& Palmer 1993;
Palmer, Lopez, Meaburn, \& Lloyd 1996) is the archetypical example of
this class of PNe. Many point-symmetric PNe exhibit the morphology
solely in  a series of opposing knots. Cliffe \ea (1995) have modeled
these systems as as episodic precessing jets.  Lopez, Vazquez \&
Rodriguez (1995) have proposed that the systems represent a separate class
of PNe defined as Bipolar Rotating Episodic Jets or {\it BRETs}. The
origin of these point symmtreic patterns is one of the most intriguing aspects of PNe
studies. As we discussed in section 2.7 it would seem that models which
invoke magnetic fields near the source may provide a better means for
generating point symmetry in jets than hydrodynamic models.  This
point needs to be explored further. There are also a number of
bipolar PNe which show a point-symmetric brightness distribution imposed
on their opposing lobes even though no jet is visible (Hb 5, Corradi \&
Schwarz 1993c). These systems are
particularly puzzling and their origin remains relatively unexplored.
Cliffe \ea (1995) conjectured that the bipolar lobes in these systems
represented {\it global} bow shocks surrounding precessing jets.  This
raises the issue that PNe jets may not always appear as the bright
central beams like those associated with YSOs.  There are highly
collimated PNe (M2-9; Balick, Icke \& Mellema 1998) and PPNe
(Bobrowsky \ea 1998, Trammell \& Goodrich 1996) which appear hollow.  Why
only some PNe would produce visible jets is unknown.

Inspection of the catalog of Schwarz Corridi \& Melnick 1992 shows that
in total at least 7\%  of all PNe exhibit either the {\it S} shaped jets or
the point symmetric brightness pattern. In addition other classes of
bipolar outflows (LBVs; Weis \ea 1996, YSOs; Reipurth, Bally \& Devine 1997)
show point symmetry making the issue rather pressing for future
studies.

\subsection{Luminous Blue Variables}

Luminous Blue Variables (LBVs) are massive unstable stars known to
experience mass ejections and dramatic increases in luminosity (Humphreys
\& Davidson 1994).  Recent observations have revealed a number of
LBVs or LBV candidates to be surrounded by extended aspherical outflows.
The most extraordinary of these is the markedly bipolar nebula
surrounding $\eta$ Carinae (``the homunculus'': Hester \ea 1991 Ebbets
\ea 1993 Humphreys \& Davidson 1994).  Other LBVs show nebulae with
varying degrees of asphericity from elliptical (R127) to strongly
bipolar (HR Carinae: Nota \ea 1995; Weis \ea 1996).

These shapes are quite similar to what has been observed in Planetary
Nebulae (PNe) leading to the suggestion that both families of objects
are shaped in similar ways. So far the homunculus of $\eta$ Car is the
best studied of all LBV bipolar outflows. $\eta$ Car is an extraordinary
star, one of the massive known in the galaxy $M \approx 100$ \msol.
 In the 1840s
it became the second brightest star in the sky. This {\it outburst}
must have been accompanied by a rapid and dramatic increase in mass
loss.  Proper motion studies of the Homunculus (Currie \ea 1996a)
indicate that the bipolar nebula was expelled during the giant outburst.
One of the most suggestive features in early HST images of the
homunculus was an apparent disk of material surrounding the waist of the
bipolar lobes.  This configuration has been described as an ``ant in a
tutu'' by Davidson (private communication 1995). The presence of a disk
naturally suggested that the bubble was formed via the GWBB process. In
Frank, Balick \& Davidson (1995 hereafter: FBD) a GWBB model for $\eta$
Car was presented in which a spherical {\it outburst wind} expelled
during the 1840 outburst expanded into a toroidal (disk-shaped) {\it
pre-outburst wind}.  FBD showed that the resulting bipolar outflow could
recover both the gross morphology and kinematics of the Homunculus.

Nota, Livio Clampin \& Schulte-ladbeck (1995 hereafter NLCS) presented
one of the first catalogs of LBV nebulae. Using a model similar to the
one invoked by FBD, they presented a unified picture of the development
of LBV outflows.

Garcia-Segura, Langer \& MacLow 1997 (hereafter GLM) presented a model
for $\eta$ Car which also relied on the GWBB scenario but changed the
order of importance of the winds. The novel aspect of GLM's study was to
include the effects of stellar rotation. Using the WCD model for the
first time in a bipolar bubble calculation, GLM showed that a strong
equator to pole density contrast would likely form {\it during the
outburst} when a star is close to the Eddington limit, 
$\Gamma = L/L_{Edd} \approx 1$.  Since the
critical rotation speed goes as $\sqrt{\Gamma - 1}$ stars close to their
Eddington luminosity do not have to be spun up to produce a wind
compressed disk. Thus in the GLM model the outburst wind creates the
slow torus and the {\it post-outburst} mass loss (which was not
considered in either FBD or NLCS) acts as the fast wind inflating the
bipolar bubble. GLM's models also included initial seed perturbations
which drove thin shell instabilities (Vishniac 1983, 1994). These
fragmented the shell in ways which may, in a fully 3-D model, lead to
small scale structures similar to the ``worms'', corrugations and dark
lanes seen on the surface of the $\eta$ Car's bipolar lobes (Morse \ea
1998).

As more detailed images and analysis of the homunculus have become
available (Morse \ea 1998) it has become clear that disk surrounding $\eta$
Car is not cylindrically symmetric.  Rather than a continuous structure
it appears to be a discontinuous clumpy "skirt" of debris.  The large
angular gaps in the skirt make any GWBB model which invokes an extended
disk difficult to support. It is hard to image how such a structure
could simultaneously confine a spherical outflow and become highly
fractured on large angular scales.  In addition Morse \ea 1998 and
Currie \ea 1996b have shown that the morphology of the lobes is more like
that of two flasks connected at their necks rather than two osculating
bubbles. We note that AG Car shows a similar morphology (NCLS).

To rectify this situation and embrace the full range of LBV nebulae
Frank \ea 1998 have performed simulations which further generalize
the GWBB scenario. These models turn the GWBB scenario on its head by
considering the case of an {\it aspherical fast wind} interacting with a
{\it isotropic slow wind}.  Frank \ea 1998 imagined a fast wind ejected
with higher velocity along the poles than along the equator.  There is
both observational and theoretical support for this idea. Lamers \&
Pauldrach 1991 have demonstrated that rotation induced changes in optical
depth can produce a bistable wind with high velocity and low density in
the poles and low velocity and high density in the equator.  
Observations of the wind of AG Carinae (Leitherer \ea 1994) imply a
pattern of densities and velocities from pole to equator much like that
described in Lamers \& Pauldrach 1991.  Owocki \ea 1996 have also shown
that fast poleward directed winds may be a natural consequence of hot
rapidly rotating stars.

The simulations of Frank \ea demonstrated that bipolar outflows which
matched the shapes of LBV nebulae could be produced via aspherical fast
winds. Fig 3.3 shows a comparison of the models of Frank \ea and MCL
with the LBVs $eta$ Car and AG Car.  Note that the aspherical fast winds
appear to produce a flask morphology better than the models involving a
WCD. Clearly one drawback of the Frank \ea scenario is the lack of an
equatorial spray in a nebulae like $\eta$ Car.  Recent evidence however
suggestes that the equatorial spray was formed during a later smaller
outburst in 1890 (Smith \& Gertz 1998).

More recently Dwarkadas \& Balick 1998 have attempted to deal
with the issue of eta Car's equatorial region by assuming the existence
of dense ring of gas on scales of $R = 10^{15}$ cm. The ring is confined
to a narrow region in the equator so that ($\delta R/R < 1$). These
models recover the flask shape with out an extended disk.  In
addition as the ring is accelerated outward it fragments due to
dynamical instabilities (Jones, Kang, \& Tregillis 1994). Since it is
easier to destroy a thin ring via non-axisymmetric instabilities than an
extended disk this model holds the promise of of creating both the flask
shaped bipolar lobes and the equatorial spray.  Of course the issue then
becomes the origin of the ring.

We have not considered Wolf-Rayet nebula extensively in this review
because they are generally spherical or ellipitical rather than bipolar
There are exceptions however (NGC 6888, Garcia-Segura \& MacLow 1995).
 This raises the question: why are there no strongly bipolar WR
bubbles? The answer may lie in the evolutionary considerations.  A
connection between WR stars and LBVs has been proposed by Langer \ea
1994 and the evolution of the star and the wind has been linked to
instabilities in WBB shells (Garcia-Sequra, Mac Low \& Langer 1996). 
It is possible that WR stars are too far from the Eddington
limit for rotation to induce the WCD mechanism.  If the star evolves
into a near Eddington-limit LBV an equatorial density enhancement may
form which then will lead to a bipolar bubble.  This is an attractive
idea however it begs the question of why so many PNe exhibit bipolar
shapes when they are not close to their Eddington luminosities.
Further research is needed on this point and currently the lack of
bipolarity in WR bubbles remains a mystery.

\subsection{Supernova 1987A}

The rings surrounding SN87A (Burrows \ea 1995) provide yet
another example of a circumstellar bipolar outflow.  The intense
scrutiny applied to SN87A make it a unique laboratory for studying the
connection between bipolar outflow and stellar evolution.  
Recently Brander \ea
1997 have discovered a bipolar outflow surrounding Sher 25 a star is
similar to the progenitor of SN87A.  Thus it is possible that SN87A
defines a new class of bipolar outflows.

When the central ring of SN87A was discovered it was quickly
interpreted as the waist of a bipolar WBB (Luo \& McCray 1991, Wang \&
Mazzali 1992). In these models the ambient medium is taken to be a
toroidal shaped wind deposited by SN87A's progenitor in a Red Supergiant
(RSG) stage. The bubble was subsequently inflated by a fast wind from the
star's penultimate incarnation as a Blue Supergiant (BSG). The first
numerical simulations of this process were carried out by Blondin \&
Lundqvsit 1993) who used an inverted form of the Icke function (equation \ref{ad2})
for the slow RSG wind.  Blondin \& Lundqusit 1994 were quite successful
in demonstrating the feasibility of the GWBB paradigm for SN87A. The low
expansion speed of the central ring ($V \approx 8$ \kms) presented a
problem however.  To create a model with the correct kinematics, Blondin \&
Lundqusit were forced to adopt rather low values of both the RSG and BSG
wind wind velocities. Their value of the RSG wind speed, $5$ \kms ~seemed
particularly anomalous.  Canonical values for the RSG wind speeds is
of order 20 km/s. Blondin \& Lundqusit were also
forced to take an equator to pole density contrast of $q = 20$ which at
the time seemed large.   The size of $q$ was one reason cited by McCray
and Lin 1992 in their arguments that the ambient density distribution
represented a remnant protostellar disk rather than a stellar wind.

The discovery of additional upper and lower rings of SN87A both
confirmed and confused the image of SN87A's nebula as a bipolar outflow.
Initially the extra rings were interpreted as traces of a precessing
jet (Burrows \ea 1995).  Martin and Arnett (1994) carried out
GWBB simulations similar to Blondin and Lundqvist and claimed that the
upper and lower rings were limb brightened projection of the bipolar
lobes. Martin and Arnett's models were noteworthy in that they included
the transport of radiation from the supernova and followed the
concomitant microphysics (ionization, etc.) in the nebula.  They confirmed
Blondin \& Lundqusit's results including the need for low RSG wind speeds $<
10$ km/s. While their conclusion concerning the upper and lower rings
was based on the production of synthetic images (Fig 3.4) it does not mesh
well with the detials of the observations.

Fig 3.4 shows the star at the center of the two nested ellipses formed
by the projection of the lobes. This is identical to what was seen the
models of Frank \ea 1993 (Fig 3.1). There is, however, a striking lack of
symmetry in the upper and lower rings of SN87A not present in the
synthetic image. The real rings are
considerably off center with respect to the star. Martin and Arnett were
cognizant of these issues and suggested that motion of the nebulae
through the ambient medium might distort the lobes in and beak the
symmetry of the rings. This conjecture will need to be tested using 3-D
models. Consideration of the HST images also shows that in addition to
the symmetry breaking the rings do not exactly close on each themselves.
Finally it is not clear if the actual intensity structure of the rings
can be identified with limb brightening of a lobe. Thus is appears
that the rings are a real density or emissivity structure
inherent to the lobes and a number of authors have developed
models for their formation

Dwarkadas \& Chevalier 1995 considered the effect of the UV flux from
the Blue Supergiant proposing that the upper and lower rings are the
location of an ionization front breakout.  Photons escape from the low
density polar sector of the lobes but are trapped at lower latitudes.
The rings have also been explained via pre-shaping of the RSG envelope
via ionization (Meyer 1997) or the action of a binary companion
(Podsiadlowski, \ea 1991; Lloyd, O'Obrien \& Kahn 1995). None of these 
scenarios has proven more compelling than the others and the upper and
lower rings remain one of the fundamental mysteries of SN87A.
Their nature may become more apparent when the supernova blast wave overtakes the
central ring and illuminates the rest of the nebula (Luo \ea 1994).

The existence of the aspherical RSG wind has led a number of authors to
posit the existence of a binary companion for the progenitor of SN87A.
Using arguments from stellar evolution theory along with the need for a
high equator to pole contrast, Podsiadlowski (1992) has calculated a likely
progenitor mass of 3 to 6 \msol with the progenitor taking values of
16 \msol.  In a recent calculation Collins \ea 1998 attempted
to form a direct link between the shape of the bipolar WBB and the
binary status of progenitor. Using the WCD model to specify the shape of
the RSG wind and the Guiliani solution to calculate the bubble evolution
Collins \ea demonstrated that a wind compressed environment could
reproduce correct shape of SN87A's nebula including the kinematics of
the inner ring. A RSG wind of $20$ km/s could be used since the WCD
model naturally produces low velocities in the equator.
The rotation speeds needed by the WCD model were too high ($\Omega > .3
\Omega_{c}$) to be explained by single star models.  Thus a companion
was required to spin up the star. Using a simplified analysis of the
merger of the progenitor with mass $M_p$ and a companion with mass $M_c$ 
seperated initially by a distance $a$, Collins \ea used
conservation of angular momentum to derive the following equation 
\begin{equation}
\sqrt{{G (M_p M_c)^2 \over M_p + M_c} a} = \zeta M_e R_e^2 \bar{\Omega}.
\end{equation}
\noindent $\zeta$, $M_e$ and $R_e$ correspond to the moment of inertia,
mass and radius of the envelope of the merged star respectively and 
$\bar{\Omega} = 
\Omega / \Omega_{c}$. Solving the above for the mass ratio of the two stars allowed
Collins \ea to set a lower
limit on the companion mass of $M_c ~ .4 - .7$.  While the actual mass
of a companion is likely to be larger than this, the low mass predicted
is consistent with the lack of an observed binary in SN87A 
(Crotts, Kunkle \& Heathcote 1995).

\subsection{Young Stellar Objects}

As stars evolve to the main sequence they experience vigorous episodes
of mass loss. Observations have shown that outflows from YSOs take two
striking and fairly ubiquitous forms.  Molecular outflows define the
first class. These are large-scale bipolar flows of material seen
primarily in the lines of molecules such as $CO$ and  $NH_3$.
Typical speeds of the outflows are on the order of $10$ km/s. The second
generic class of outflow are highly collimated jets, visible primarily, in
the optical.  The jets are also bipolar in the sense that opposing
``counter jets'' are often seen. These jets have proper motions which
indicate velocities of order $\ge 200$ \kms. Both the jets and
outflows show features which imply variations in the outflow direction 
and velocity.  YSO jets and molecular outflows have been the subject
of intense research activity and a number of excellent review articles
exist (Bachiller 1996, Reipurth 1997).  The connection between these objects
and other classes of bipolar stellar outflow has not been explored
however.  In this section we will briefly review two issues concerning
YSOs which are relevant to bipolar outflows as a broad class of
phenomena: (1) jet collimation; (2) the origin of molecular outflows.

{\bf Jet Collimation} As was discussed in previous sections the current
consensus holds that YSO jets are launched and collimated by MHD
processes.  The most popular models rely on magneto-centrifugal forces
in either an accretion disk (``Disk-winds''; K\"onigl 1989, Pudritz 1991)
or at the disk-star boundary (``X-winds''; Shu 1997).  These studies
have been quite successful in articulating the physical properties of
MHD collimation processes. Indeed, numerical simulations Ouyed \&
Pudritz 1997 have recently demonstrated the ability of disk-wind models
to produce both steady and time dependent jets.  Models which rely on
the interaction of a dipole stellar field tied to an accretion disk have
also shown promise (Lovelace, Romanova, Bisnovatyi-Kogan 1995, Goodson
Winglee \& Bohm 1997).

In spite of these successes many questions remain.  The existence of
a variety of working models makes it uncertain which
mechanism YSOs choose if MHD is the dominant launching {\it and}
collimation process.  There are also questions as to the effectiveness
of {\it collimation} in MHD models.  Shu \ea 1995  point out MHD
collimation can be a slow process occurring logarithmically with height
above the disk.  Ostriker 1997 has pointed out self-similar MHD disk winds
have low asymptotic speeds in cases where they become fully
(cylindrically) collimated with $B$ decreasing faster than $1/R$ with
radius.  In addition, a different set of numerical simulations (Romanova \ea
1997) found that while the magneto-centrifugal process was effective at
launching a wind, it did not produce strong collimation of the wind into
a jet.

Along with these issues, the pure hydrodynamic SFIC and converging
conical flow mechanisms discussed in section 2.4 and 2.5 can be
surprisingly effective at producing jets in YSO environments.
Hydrodynamic collimation in YSOs is not a new subject.  K\"onigl (1982)
presented a very complete study of deLaval nozzles as a means for
producing collimation in YSO outflows. Raga \& Cant\'o (1989) explored the
role of cooling in deLaval nozzles appropriate to YSOs.  The original
work on conical converging flows by Cant\'o and collaborators was directed
towards YSOs. In many of these studies a stellar wind interacting with
environmental {\it pressure} gradients produced  a stationary aspherical
wind blown cavity.  The strongest objection to these models 
was the collimation scale length (K\"onigl \& Ruden
1993) .  Observations show that in many
jets collimation must occur on scales of order $R \approx 10 AU$ (Ray
\ea 1996).  This is too small for pressure
confinement.  The more recent studies of Frank \& Mellema (1996) and
Mellema \& Frank (1997) which applied time-dependent WBB hydrodynamic
collimation mechanisms to YSOs attempted to address these issues. These
papers demonstrated that the SFIC and converging conical flow processes
could generate well collimated supersonic jets in environments
appropriate to the earliest or ``Class 0'' protostars.  In Mellema \&
Frank (1997) the effects of both inertial and ram pressure confinement
from the inflowing cloud where also considered.  Using a simple model
for the evolution of a bubble driven by a {\it time-dependent wind} it
was demonstrated that infall ram pressure created a bubble which
oscillated or collapsed back onto the star. In this way WBBs could
hydrodynamically collimate jets on the correct scales.  Wilkin and Stahler
(1998) have applied a quasi-static shell model to hydrodynamic
YSO collimation.  They find the changes in the geometry of the enviroment
will also produce a bipolar outflow. Their timescales for 
bubble evolution were too long however and they also suggested 
oscillation via time-dependence in the wind.

We note that Frank \& Mellema separated the issues of launching a wind
from collimating it into a jet. The luminosity associated with low mass
stars is too small for produce a radiation driven outflow. Thus the
winds from low mass protostars probably require MHD processes.  The
collimation of the wind into a narrow jet, however, may require
interaction with the environment.  For high mass protostars there is no
need to invoke a strong magnetic field to either launch or collimate
observed jets/outflows (Shepherd \& Churchwell 1996, Churchwell 1997).
The strong UV flux from massive stars may also play an important role
in driving hydrodynamic collimation.  Yorke \& Weltz 1996 and Richling
\& York 1997 have shown that hot material flowing off a ionized
accretion disk can effectively constrain a stellar wind. The gas
streams off the disk in those regions where the escape speed is
comparable to the sound speed. These {\it thermal disk winds} can
constrain the stellar wind and lead to aspherical wind shocks. Flow
focusing of post-shock wind material into a jet occurs in a manner
similar to SFIC process.

It is likely that MHD processes play a significant role in producing
launching many YSO jets. The results presented above however show that
the environment needs be considered in models of the collimation
process spanning the different evolutionary stages and environmental
conditions associated with star formation.  The effect of the environment 
can also be inferred from observations.  Some jets systems show collimation
increasing with distance from the source Ray \ea (1996), an indication
the environment is helping narrow the jet as it propagates. The
existence of cavities at the base of some jets has also been observed
suggesting the presence of a wider outflow on smaller scales (Close
1997, Cabrit \ea 1996). The shocks required for hydrodynamic
collimation may also be inferred from ionization fractions in jets
which tend increase close to the protostellar source (Bacciotti 1997).
Evidence of strong non-thermal emission due to shocks exists close to
the source of at least one jet system (Reid \ea 1995, Wilner \ea
1997).

{\bf Driving Molecular Outflows} The second issue connecting YSO
phenomena to other bipolar outflows is the Jet/Molecular outflow
connection (Cabit \ea 1997). After almost two decades of study it is still
unclear if the YSO jets and molecular outflows are causally related or
simply co-extensive.  Competing models invoke either ``wide-angle wind''
or a jet to drive the molecular outflows.

GWBB scenarios for the origin of the
molecular outflows have been invoked since their discovery
(Snell, Loren \& Plambeck 1980,
K\"onigl 1992).  In Shu \ea 1991 a momentum conserving interaction
between a gaseous torus and an aspherical central wind was explored.
Given a description for the asphericity of the environment,
$P(\cos\theta)$, and the wind, $Q(\cos\theta)$,  the Shu \ea model allows
the shape and speed of the outflow to be simply characterized:

\begin{equation} {d R_{as} \over d t} = V_{as} 
= ({\dot{M}_w c_s V_w \over \dot{M}_a})^{1/2}
({P(cos\theta)\over Q(\cos\theta)})^{1/2}.
\end{equation}

\noindent where $M_a$ is the accretion rate and $c_s$ is the sound speed in the
infalling gas. Fig 3.5 shows the shape of a WBB
formed when a spherical wind expands into a collapsing filament
(Delemarter, Frank \& Hartmann 1998, Hartmann \ea 1996).

The alternative sceario holds that molecular outflows are driven not by winds
but by fully formed jets.  Masson \& Chernin 1993 explored the ability
of straight constant velocity jets to drive molecular outflows via
``prompt entrainment'' at a bow shock.  ``Steady entrainment'' jet
driven models where outflows result from viscous boundary layers at the
jet/ambient-medium interface have also been explored (Raga {\it et al.}
1993; Stahler 1993).

There are two characteristics of Molecular Outflows that are
particularly vexing for either theoretical scenario to explain
satisfactorily.

1) The degree of collimation varies among different objects. Some
outflows have been observed to be highly collimated, with length to
width ratios of as great as 20:1. Others are very wide, with this ratio
near unity.

2) The momentum in the outflows is primarily {\it forward driven}. This
means most of the velocity vectors in the flow are oriented along the
long axis of the lobes.

While GWBB models can produce wide outflow lobes they have difficulty
producing the correct momentum distribution in the lobes.  If the
molecular outflows are formed as ``energy conserving'' WBBs the
forward-driven characterstics would be difficult to achieve (Masson \&
Chernin 1993).  In such a case the lobes would be inflated by the
pressure of shocked stellar wind material.  Since thermal pressure
always acts normal to the surface of the lobe significant velocities
transverse to the axis of the lobe should be observed. Observationally
these transverse motions would appear as both red and blue shifted
velocity components from {\it each} lobe of a bipolar outflow.  Studies
of molecular outflows, however, have not revealed the presence of
transverse motions.  In general, blue (red) shifted material dominates
in the blue (red) shifted bipolar lobe. A recent study by Lada \& Fich
(1995) emphasized this point as their observations of NGC 2064G reveal
20:1 ratios of blue to red-shifted gas in the blue lobe. Even momentum
conserving models Shu \ea 1991 would have difficulty producing strong
forward driven flows.

Masson \& Chernin 1992 examined the observational consequences of the
Shu \ea 1992 momentum conserving WBB model. They found that unless
extreme distributions for $P(\theta)$ and $Q(\theta)$ were adopted,
the model failed to recover the line profiles and momentum
distributions observed in molecular outflows.  On the other hand
calculations by Li \& Shu (1996) have shown that magnetized clouds can
collapse to form dense toroids providing very steep $P(\theta)$
distributions.  The ``wide angle'' flows produced by the X-wind model
also producestrongly focused $Q(\theta)$ distributions. Together
the X-wind and collapsed magnetized clouds models may make the GWBB scenario
more effective.

Jet driven outflow models of molecular outflows have thier own
strengths and weaknesses.  Jets can produce the correct "forward
driven" momentum distribution better than wide angle winds however they
have difficulty explaining the different degrees of outflow
collimation.  Masson \& Chernin (1993) and Chernin \& Masson (1995) 
have suggested that
wandering or precessing jets may produce outflows which fit the
observational constraints.  In the wandering jet model the jet head
drives different parts of the ambient cloud as the propagation
direction changes.  Both Cliffe, Frank \& Jones (1996) and Suttner \ea
(1997) have demonstrated that precessing jets will in fact produce
a wide global bow shock enveloping the entire ``corkscrew'' of the jet
(Biro \& Raga 1994). If this global bow shock structure can be
identified with molecular outflows it may allow jet driven models to
recover the spectrum of outflow shapes as the result of different 
``wandering'' angles.  Velocity variations (Raga \&
Cabrit 1993) in the beam have also been used as a means for widening
jet driven outflows.  In these models the bow shock is inflated via
pressure from gas``squirted'' out the sides of internal shocks. 

Finally it should be noted that the Shu \ea 1994 X wind model is
distinctive in that the jets are actually an optical illusion due to
density stratification in a wide angle winds.  Such a model has
advantages when considering the need to create both jets and molecular
outflows.

\section{Summary} 
The studies reviewed in this paper demonstrate
bipolar outflow formation is a common process associated with stellar
evolution. Moreover, the similarieties between structural components of
the outflows from low and high mass, evolved and young stars would seem
to imply that similar processes shape them all. Such an implication is
still open to debate however and it is useful to briefly consider what
outflows at different ends of the evolution and mass spectrum have in
common.

Bipolar outflows from evolved stars share many
properties.  Wide-lobed bipolar Planetary Nebulae (PNe), Luminous
Blue Variables (LBVs) nebulae and Supernova 1987A are quite
similar in both their kinematics and shape.  Given what is already
understood about mass loss from these stars the Generalized Wind
Blown Bubble paradigm seems an approprate approach for describing the 
formation of bipolar outflows in many these stars. The existence
of non-axisymmetric features in some bubbles however argues that 
this model requires either additional inputs (ionization effects, clumpy flows 
Hartquist \& Dyson 1996, etc.) or a move to MHD scearios.

The relation between bipolar outflows in evolved stars and Young
Stellar Objects is more complicated.  In YSOs the outflows are most
likely associated with accretion disks. While low mass stars would not
be able to produce any form of outflow without a strong magnetic field,
high mass
YSOs may not require magnetic fields to launch or collimate winds.
The presence of long narrow jets {\it and} bipolar outflows presents a
twin challenge for theories which might link evolved and young stellar
systems.  As we have seen GWBB models can produce both extended
outflows and narrow jets but it is not clear if it can produce both
simultaneously.

Recent studies of PNe and Proto-PNe revealing the presence of jets in
evolved stars systems offer the hope of finding links betweeb
evolved staroutflows and YSOs.  PNe jets often bear strinking resembelnce to
those in YSO often appearing as smaller, shorter versions HH systems and
displaying similar point-symmetric shapes.  For some of these systems
hydrodynamic collimation models may work quite well (Dopita 1996). In
others the Magnetized Wind Bubble scenario may be more attractive
alternative.  The suggestion that accreation disks form in binary AGB
systems (Livio \& Soker 1994) provides yet another potential link between
evolved and YSO bipolar outflows.

Along with physical processes different classes of bipolar outflows are
connected by the tools and methodology used to study them.  Both
observational diagnostics/techniques and numerical models can often be
readily adapted between different classes of outflow. In some cases the types
of questions that can be asked are identical.  For example the study of
LBV nebula has led to questions about the geometry of the fast stellar
wind which can be answered by examination of the mass and momentum
distributions in the bipolar lobes. Such a procedure was used by Masson
\& Chernin 1992 in their examined of the WBB model of YSO molecular
outflows.

\noindent{\bf Stellar Wind Palentology} 

Given their ubiquity what do bipolar outflows teach us about stellar
evolution?  One of the most compelling reasons for studying 
outflows is the stellar history encoded in their structure. The
outflows trace the history of mass loss in the star. For systems
where the succesive wind phases form extended circumstellar evelopes,
the bipolar outflows can provide clues to evolutionary
changes in $\dot{M}$ and $V_w$.  For some systems the
structure of the outflows can reveal properties of the pre-existing
environment.  Thus the outflows can teach us about the history and physics
of the stars themselves.

The high quality of observational data available combined with the variety of 
``lookback'' timescales ($t_d$, equation \ref{bas}) inherent
to the outflows  ($t_d \sim 10^2 -  10^5 ~y$) offers the possibilty that
key transitions in an individual star's history might be recovered if
we learn where and how to look. The use of collimated outflows as fossils
can succeed if researchers can identify issues which provide a
reasonably unique and unambiguous bridges between nebular
stucture/dynamics and stellar evolution.

The development of  nebular palentological studies
is still in its infancy. Researchers have, however, learned enough that
one can point to examples of what such a investigation might look like.
SN87A and $\eta$ Carinae are both unique objects.  In each case 
the intense scrutiny the outflows have received has allowed their
histories and the history of the central stars to be specfied in
some detail.  There are so many PNe and YSOs that a similar
critical mass of data has not accumulated around many of these objects as
of yet. 

The nebula surrounding SN87A is a particularly good example to consider.
The properties of the central ring are known with
enough accuracy that GWBB models can place strong constraints on the
mass loss rates of both the Blue Supergiant and Red Supergaint winds.
Assuming that the upper and lower rings are associated with the bipolar
lobes then the geometry of Red Supergaint wind and the equator to pole
density contrast can also be constrained.  This, in turn, leads to 
some limitation on the stellar rotation rate and the
potential properties of a binary companion.  New studies of the
extended circumstellar environment reveal a mild aspherical geometry on
with lookback timescales of $t_d \sim 10^5 ~y$ (Crotts \& Heathcote 1997).  
It is possible that these data may provide futher evidence of evolution of rotation
and binary interactions.  While many of the conclusions rely on the
application of specfic models (GWBB, Wind Compressed Disks),
as research progresses one would
expect alternative models to be explored with as much detail.  

The studies of $\eta$ Carinae have also allowed the history of the
enigmatic central star to be sketched out.  The expansion patterns of
the lobes clearly link them to the outburst of 1843.  The attention
focused on the equatorial ejecta however allows very different models
for the history and geometry of the pre-outburst, outburst, and
post-outburst winds to examined in detail.  It should also be possible to
link the strutures which mottle the surface of the Homunculus' to
instabilites dependant upon the the form of $\dot{M}(t)$ and $V_w(t)$

With these examples it is not unresonable to expect that progress in
bipolar outflows studies will someday allow the evolution of
fundemental stellar properties to be read off the outflows.  The effect
of mass loss, rotation, magnetic fields, and binary companions on
stellar birth and death are all questions of fundemental importance.
The rapid progress being made in bipolar outflow studies will produce
insights not only into hydrodynamic and hydromagnetic phenomena in
general but also provide direct links from the physics of these nebulae
to the properties of the stellar sources.  Taken together these
prospects will make the study of bipolar outflows an exciting field for
years to come.

\noindent{\bf Acknowledgments} 
There are so many people who have
contributed in to this review through interesting and helpful
discussions that I do not have space here to acknowledge them all. I
would however like to thank Bruce Balick, Garrelt Mellema, Vincent
Icke, Mario Livio, Tom Jones, Dongsu Ryu, Jack Thomas, John Dyson,
Mordecai-Mark Maclow, Guy Delamarter and Tom Gardner for their help.
Support for this work was provided at the University of Rochester by
NSF grant AST-9702484 and the Laboratory for Laser Energetics.

\begin{center}
{\bf FIGURE CAPTIONS}
\end{center}
\begin{description}

\item[Fig.~1.1]{Bipolar Outflows across the HR diagram.  Four bipolar
outflows from different classes of star.  Upper Reft: the Luminious Blue
Variable $\eta$ Carinae (Morse \ea 1998). Upper Right: the planetary 
nebula M2-9 (Balick, Icke \& Mellema 1998). Lower left:
the nebula surrounding Supernova 1987A (Burrows \ea 1995). Lower Right:
the red shifted (top) and blue shifted (lobes) of NGC 2264G (Lada \& Fich 1995)}

\item[Fig.~1.2]{The Shapes of Bipolar Outflows (PNe).  Six planetary nebulae
exhibiting a range of morphologies. From left to right, top to bottom:
IC 3568, round (Bond \& Ciardullo 1998); NGC 6826, elliptical
with ansae (Balick \ea 1997), NGC 3918, bipolar with jets 
(Bond \& Ciardullo 1998); Hubble 5, bipolar (Balick, Icke \& Mellema 1998),
NGC 7009 elliptical with jets/ansae (Balick \ea 1997); NGC 5307,
point-symmetric (Bond \& Ciardullo 1998)}

\item[Fig.~2.1]{Schematic of the 3 types of Wind Blown Bubble.  
The ambient shock $R_{as}$ and wind shock $R_{ws}$ are labeled.
The contact discontinuity (CD) seperating the two shocks is not.}

\item[Fig.~2.2]{Schematic of the Generalized Wind Blown Bubble scenario 
(Icke 1996). A fast tenuous wind (arrows) from the central star
expands into a ambient medium with an aspherical (toroidal) density
distribution. The shape of the contact discontinuity (CD) indicates 
it is prone to instabilities. Note also the prolate shape of the wind shock
(see section 2.4).

\item[Fig.~2.3] {Evolution of 4 GWBB planetary nebula models.  Shown
are greyscale $Log_{10}$ density maps at six different times for each
model.  The environment corresponds to the Icke function (eqn
\ref{ad1}).  The pole to equator contrast $q = \rho(90^o)/\rho(0^o)$
increases from top to the bottom of the figure.  Light scales correspond
to high density. The ambient shock and swept-up shell appear as the
light grayscales surrounding the bubble.  The wind shock appears as the
interface between discontiniuity in light to dark grayscales.  Note the
increasing ellipticity of the inner shock as one moves down the figure.
These models were computed with a radiation-gasdynamic code and they
show that the GWBB model can produce a wide array of bubble shapes.
See Frank \& Mellema 1994b for details.}

\item[Fig.~2.4] {Comparison of numerical and analytical GWBB models.  
Solid lines show the results of self-similar calculation for the shape
of the ambient shock $R_{as}(\theta)$. Dashed-lines show the ambient shock
position taken from numerial models with with identical initial conditions
(Dwarkadas, Chevalier \& Blondin 1996)}

\item[Fig.~2.5] {Density, velocity and temperature for a SFIC jet.
Shown are a gray scale map of $\log_{\rm 10} (\rho)$, a vector map of
velocity and a $\log_{\rm 10} (T)$. Dark (light) shades correspond to
low (high) densities.  In the velocity field map vectors in the inner,
freely expanding wind zone have not been plotted.  Thus the first
``shell'' of vectors maps out the wind shock. The dark solid
line in the map of T marks the contact discontinuity.  Note the extreme
aspherical shape of the inner shock. Note also the internal shocks in the
body of the jet.  For details see Mellema \& Frank 1997.} 

\item[Fig.~2.6] {Schematic for Jet Collimation via conical converging
flows.  A prolate momentum conserving bubble develops via a wind environment
interaction.  Shocked wind material forms a thin shell and is forced to stream 
along the CD until it forms a conical converging flow at the pole. A new shock
forms which redirects material into a jet.  See Frank, Balick \& Livio
1996 for details.}

\item[Fig.~2.7] {Left: Protoplanetary nebula He 3-1475 in [N II] 6584 filter.
Right:  Numerical simulation showing
hydrodynamic collimation of a converging conical flow.  Dark regions
correspond to the high density.  Dashed line outlines the
position of the collimating shock in an analytic model (Borkowski,
Blondin \& Harrington 1997)}

\item[Fig.~2.8] {Magnetized Wind Bubble. Bottom to top: Density, 
Magnetic Pressure ($P_b = B_\phi^2/8\pi$), and
Total Pressure, ($P_{tot} = P_b + P_g$) for a magnetized wind bubble.
 The velocity field is superimposed on
the $P_{tot}$ contours. (Rozyczka \& Franco 1996)}

\item[Fig.~2.9] {3-D Magnetized Wind Bubbles from Precessing Star
(Garcia-Segura 1997). 4 models: A (top right), two-wind
calculation (t=900 yr). B (top left), three-wind
calculation (t=2750 yr). C (bottom right), 
two-wind calculation including  precession (t=900 yr). 
D (bottom left), three-wind calculation including
precession (t=1300 yr). A and C show the logarithm of
the emission measure. B and D are synthetic observations.}

\item[Fig.~3.1] {Theory Confronts Reality. Left: Six synthetic $H\alpha$ images
taken from a single simulation (second from bottom Fig 2.3).  Each image is 
inclined at different angles ($i$) with respect to the line of sight.  
Right: Likely nebular counterparts. (Frank \ea 1993, Balick 1987)}

\item[Fig.~3.2] {Synthetic position velocity diagrams for PNe simlation.
Projected [OIII]$\lambda5007$ long-slit spectrum maps for model shown
in figure 3.1.  The position axis is plotted in units of $1\times10^{16}$ ~cm.
The velocity axis is plotted in units of \kms.}

\item[Fig ~3.3] {Images and GWBB models of LBVs.  LBV observations:
(top left) $\eta$ Carinae (Morse \ea 1998); (bottom left) AG Carinae 
(Nota \ea 1995). GWBB models of LBV nebulae: (top right) Spherical fast wind
expanding into aspherical slow wind formed via the WCD mechanism (Garica-Segura
\ea 1997); Aspherical fast wind expanding into spherical slow wind (Frank
\ea 1998)}

\item[Fig ~3.4] {GWBB model of SN87A.  Left: Numerical simulation of a
spherical BSG wind expanding into toroidal RSG wind.  Shown are density
contours and velocity vectors. Right: synthetic image of model
showing limb brightened lobes and central ring. (Martin \& Arnett 1996)}

\item[Fig ~3.5] {GWBB model for Molecular Outflows.  Left: Schematic of
momentum conserving model from Shu \ea 1992.  Right: Evolution of
Shu \ea 1992 outflow for spherical stellar wind expanding into
a collapsing filiment (Delamarter \& Frank 1998,
Hartmann Calvet \& Boss 1996)} 

}
\end{description}


\begin{thebibliography}{}


\bibitem 1 Aller, L.H., 1941, ApJ, 93, 236

\bibitem 1 Arthur, S.J., Henney, W. J., \& Dyson, J.E., 1996, A\&A, 313, 897

\bibitem 1 Asida \& Tuchman, Y., 1995, in Asymmetrical Planetary Nebulae,
eds A. Hrapaz \& N. Soker, (Bristol; Institute pf Physics Publishing)

\bibitem 1 Bachiller, R., 1996, ARA\&A, 34, 111

\bibitem 1 Bachiller, R., \& Gutie\'rrez, M. P., 1997, in Herbig-Haro Flows 
and the Birth of Low Mass Stars, in IAU Symposium no. 182, eds B. Reipurth \& C Bertout,
(Kluwer, Dortdrecht)

\bibitem 1 Bacciotii, F., 1997, in Herbig-Haro Flows and the Birth of Low
Mass Stars, in IAU Symposium no. 182, eds B. Reipurth \& C Bertout,
(Kluwer, Dortdrecht)

\bibitem 1 Balick, B. 1987.  AJ 94 671

\bibitem 1 Balick, B., Preston, H., Icke, V,  1987, AJ , 94, 1641

\bibitem 1 Balick, B., Perinotto, M., Maccioni, A., Terzian, Y., \& Hajian, A.,
1994, ApJ, 424, 800

\bibitem 1 Balick, B., Rugers, M., Terzian, Y. \& Chengular, J.N.,
1993, ApJ, 411, 778

\bibitem 1 Balick, B., Icke, V., \& Mellema, G., 1998, in preparation

\bibitem 1 Brandner, W, Chu, Y, Eisenhauer, F, Grebel, E., Points, S.1997, ApJ, 489, 153

\bibitem 1 Beiber, J.M., \& Rust, D.M., 1995, ApJ, 453, 911

\bibitem 1 Biro, S., Raga, A., 1994, ApJ, 434, 221

\bibitem 1 Bjorkman, J., \& Cassinelli, J., 1992, ApJ, 409, 429

\bibitem 1 Blondin, J., \& Lundqvist, P, 1993, ApJ, 405, 337

\bibitem 1 Bobrowsky, M., Sahu, K., Parthasarathy, M.\& Garcia-Lario, P.,
1998, Nature, 392, 469

\bibitem 1 Bond, H, Livio, M.,1990, ApJ, 355, 568

\bibitem 1 Borkowski, K, Blondin, J, \& Harrington, J., 1997, ApJ, 482L, 97

\bibitem 1 Bryce, M., Lopez, J., Holloway, A., \& Meaburn, J., 1997, ApJL, 
in press

\bibitem 1 Burrows, C. \ea 1995, ApJ, 452, 680

\bibitem 1 Cabrit, S., Guilloteau, S., Andre, P., Bertout, C., Montmerle, T.,
\& Schuster, K., 1996, A\&A, 305, 527
 
\bibitem 1 Cabrit, S., Raga, A., \& Gueth, F., 1997, in 
{\it Herbig-Haro Flows and the Birth of Low Mass Stars, IAU Symposium no. 182},
 eds B. Reipurth \& C Bertout,
(Kluwer, Dortdrecht)
 
\bibitem 1 Cant{\'o} J., \& Rodr{\'\i}guez L. F., 1980, ApJ, 239, 982

\bibitem 1 Cant{\'o} J., Tenorio-Tagle G., \& R{\'o}\.zyczka M., 1988, 
A\&A, 192, 287 (CTTR)

\bibitem 1 Chevalier, R., \& Luo, D., 1994, ApJ, 421, 225

\bibitem 1 Chevalier, R., \& Dwarkadas, V.,  1995, ApJ, 452, 45

\bibitem 1 Chernin, L., \& Masson, C. R., 1995, ApJ, 455, 182

\bibitem 1 Churchwell, E., 1997, ApJ, 479, 59

\bibitem 1 Cliffe, J.A., Frank, A., Livio, M., \& Jones, J, 1995, ApJ, 44, L49

\bibitem 1 Cliffe, A., Frank, A., \& Jones, T.W., 1996, MNRAS, 282, 1114

\bibitem 1 Close, L., Roddier, F., Northcott, M., Roddier, C., \& Graves, J.,
1997, ApJ, 478, 777

\bibitem 1 Corradi, R, L. M. \& Schwarz, H. E., 1993a, A\&A, 273, 247

\bibitem 1 Corradi, R, L. M. \& Schwarz, H. E., 1993b, A\&A, 268, 714  

\bibitem 1 Corradi, R, L. M. \& Schwarz, H. E., 1993c, A\&A, 269, 462  

\bibitem 1 Collins, T., Frank, A., Bjorkman, J., \& Livio, M., ApJ, submitted

\bibitem 1 Crotts, A., Kunkle, W., \& Heathcote, S., 1995, ApJ, 438, 724

\bibitem 1 Crotts, A., \& Heathcote, S., 1997, BAAS, 191, 4014

\bibitem 1 Crowley, S., in Solar System Magnetic Fields, ed E. Priest
(Reidel Dordrecht)

\bibitem 1 Currie, D., Dowling, D., Shaya, E., Hester, J, Scowen, P., Groth, E.,
Lynds, R., \& O'Neil E., 1996a, AJ, 112, 1115

\bibitem 1 Currie, D.,Dowling, D., Shaya, E ., Hester,  1996b, 
in The Role of Dust in Star Formation, ESO Workshop,
 ed. H. Kaufl (New York, Springer Verlag 1989)

\bibitem 1 Dagni, R., \& Soker, N., 1998, ApJ, 499, L83

\bibitem 1 Delemarter G., Frank, A., \& Hartmann, L., 1998, in preparation

\bibitem 1 Dere, K.P., 1996, ApJ, 472, 864

\bibitem 1 Dopita, M., 1997, ApJ, 485L, 41

\bibitem 1 Dorfi, E. A., \& Hoefner, S 1996, A\&A, 313, 60

\bibitem 1 Dyson, J.E., 1977, A\&A, 59, 161

\bibitem 1 Dyson, J.E., \& de Vries., 1972, A\&A, 20, 233

\bibitem 1 Dyson, J., \& Williams, R., 1981, `{\it The Physics of the 
Interstellar Medium}, (University Press, Manchester)

\bibitem 1 Dwarkadas, V., Chevalier, R.A., \& Blondin, J.M. 1996, ApJ ,457 773

\bibitem 1 Dwarkadas, V., \& Balick, 1998, submitted

\bibitem 1 Ebbets, D., Garner, H., White, R., Davidson, K., \& Walborn, N., 1993,
in {\it Circumstellar Meia in the Late Stages of Stellar Evolution, Proc. 34th
Herstmonceux Conf.}, (Cambridge: Cambridge Univ. Press)

\bibitem 1 Eichler, D., 1982, ApJ, 263, 571

\bibitem 1 Eulderink, F., \& Mellema, G., 1994, A\&A, 284, 654

\bibitem 1 Frank, A., Balick, B., Icke, V., \& Mellema, G., 1993, ApJ, 404, L25.

\bibitem 1 Frank, A., \& Mellema, G., 1994a, A\&A, 278, 320

\bibitem 1 Frank, A., \& Mellema, G., 1994b, ApJ, 430, 800

\bibitem 1 Frank, A., 1994, AJ, 107, 256

\bibitem 1 Frank, A., Balick, B., \& Davidson K., 1994, ApJ, 441L, 77 (FBD)

\bibitem 1 Frank A., Balick B.,  \& Livio M., 1996, ApJ, 471, L53

\bibitem 1  Frank A., 1995, AJ, 110, 2457

\bibitem 1 Frank, A., Ryu, D., \& Davidson, K., 1998, ApJ, in press

\bibitem 1 Frank, A., \& Mellema, G., 1996, ApJ, 472, 684

\bibitem 1 Garcia-Segura, G., \& Mac Low, M., 1995, ApJ, 455, 160

\bibitem 1 Garcia-segura, G., Mac Low, M., \& Langer, N., 1996, A\&A, 305, 229

\bibitem 1 Garcia-segura, G., Langer, N., \& Mac Low, M., 1997,
in  {\it Luminious Blue Variables: Massive Stars in Transition},
ed A. Nota \& J.G.L.M. Lamers, (ASP Conference Series, San Francisco)

\bibitem 1 Garcia-Segura, G., 1997, ApJ, 489L, 189

\bibitem 1 Garcia-segura, G., Langer, N.,  Rozyczka, M., Franco, J., \&
Mac Low, M., 1997, in {\it The Sixth Texas-Mexico Conference on Astrophysics:
Astrophysical Plasmas Near and Far}, Rev Mex AA, eds S. Torres-Peimbert
\& R. Dufour 

\bibitem 1 Garcia-Segura, G., Mac Low, M., Langer, N., 1996, A\&A, 305, 229

\bibitem 1 Garcia-Segura, G., Langer, N., Mac Low, M.,  1996, A\&A, 316, 133

\bibitem 1 Giuliani, J.L., 1982, ApJ, 256, 624

\bibitem 1 Goodson, A., Winglee, R., Bohm, K., 1997, ApJ, 489, 199

\bibitem 1 Gosling, J., 1996, ARA\&A, 34, 35

\bibitem 1 Gurzadyan, G, 1996, {\it The Physics and Dynamics of Planetary Nebulae}
(Springer, New York: Springer)

\bibitem 1 Hajian, A, \& Balick, B., 1998, preprint

\bibitem 1 Hajian, A, Balick, B., Terzian, Y., \& Perinotto, M., 1997,
ApJ, 487, 304

\bibitem 1 Hartquist, T. W.,  Dyson, J. E.,  1996, Ap\&SS, 245, 263

\bibitem 1 Hartigan, P., Morse, J., \& Raymond, J., 1993, ApJ, 434, 232

\bibitem 1 Hartmann L., Calvet N., \& Boss A., 1996, ApJ, 464, 387

\bibitem 1 Henney, W.J, \& Dyson, J.E., 1992, A\&A, 261, 301

\bibitem 1 Hester, J. J., Light, R. M., Westphal, J. A., Currie, D. G., \&
Groth, E. J., 1991, AJ, 102, 654 

\bibitem 1 Humphreys, R., \& Davidson, K., 1994, PASP, 106, 1025

\bibitem 1 Hollenbach, D, \& McKee, C, 1989, 342, 306

\bibitem 1 Innes, D.E., Inhester, B., Axford, W.I., \& Wilhelm, K., 1997,
Nature, 386, 811

\bibitem 1 Icke, V., 1988, A\&A, 202, 177

\bibitem 1 Icke, V., Balick, B., \& Preston, H., 1989, AJ, 97, 462

\bibitem 1 Icke, V., Balick, B., \& Frank, A., 1992, A\&A, 253, 224.

\bibitem 1 Icke, V., Mellema, G., Balick, B., Eulderink, F., \& Frank, A., 1992,
Nature, 355, 524.

\bibitem 1 Icke, V., 1994 in {\it Circumstellar Media in the Late Stages of Stellar
Evolution}, eds R. Clegg \ea (Cambridge, Cambridge Press)

\bibitem 1 Ignace, R., Cassinelli, J. P., \& Bjorkman, J. E., 1996, ApJ, 459, 671

\bibitem 1 Jones, T.W., Kang, H., \& Tregillis, I.L., 1994, ApJ, 432, 194

\bibitem 1 Kahn, F. D., 1983 in ``IAU Symposium 103: Planetary Nebulae''
ed D. Flower (Reidel, Dordrecht)

\bibitem 1 Kahn, F. D., \& West, K. A., 1985, MNRAS, 212, 837

\bibitem 1 Kompaneets, A.S., 1960, Dokl.\ Akad.\ Nauk, 130, 1001

\bibitem 1 K\"onigl, A. 1989, ApJ, 392, 622.

\bibitem 1 K\"onigl, A 1982 \apj, 261, 115

\bibitem 1 K\"onigl, A., \&  Ruden, S.P. 1993, in ''Protostars and Planets III'', ed.
E.H. Levy \& J.I. Lunine (University of Arizona Press), 641. 

\bibitem 1 Koo, B.C., \& McKee, C.F., 1992, ApJ, 388, 93.

\bibitem 1 Kwok, S., Purton, C. et al. 1978, ApJ 219, L125

\bibitem 1 Kudritzki, R.P., Pauldrach,A., \& Abbott, D.C., 1989,  A\&A, 219, 205

\bibitem 1 Lada, C. J., \& Fich, M 1996, ApJ, 459, 638

\bibitem 1 Lamers, H., \& Pauldrach, A., 1991, A\&A, 244, 5

\bibitem 1 Langer, N., Hamann, W.R., Lennon, M., Najarro, F., Pauldrach, A., \&
Puls, J., 1994, A\&A, 290, 819

\bibitem 1 Langer, N., 1997,
in  {\it Luminious Blue Variables: Massive Stars in Transition},
ed A. Nota \& J.G.L.M. Lamers, (ASP Conference Series, San Francisco)

\bibitem 1 Latter, W B., Kelly, D. M., Hora, J. L., \& Deutsch, L. K, 1995, ApJS, 100, 159

\bibitem 1 Lazarian, A., \& Vishniac, E.T., 1998, preprint

\bibitem 1 Leitherer, C., Allen, R., Altner, B., Damineli, A., 
Drissen, L., Idiart, T., Lupie, O., Nota, A., Schmutz, W., \& Shore, S., 1994, ApJ, 428, 292

\bibitem 1 Li Z., \& Shu F., 1996, ApJ, 472, 211

\bibitem 1 Lada C.J., \& Fich, M., 1996, ApJ, 459, 638.

\bibitem 1 Livio, M., 1994 in {\it Circumstellar Media in the Late Stages of Stellar
Evolution}, eds R. Clegg \ea (Cambridge, Cambridge Press)

\bibitem 1 Livio, M., \& Soker N., 1988, ApJ, 329, 764

\bibitem 1 Livio, M., 1997, Space Science Reviews, 82, 389

\bibitem 1 Livio, M., \& Pringle, J. E., 1997, ApJ, 486, 835

\bibitem 1 Lloyd, H. M., O'Brien, T. J., \& Kahn, F. D.,  1995, MNRAS, 273, 19

\bibitem 1 L\'opez, J.A., Meaburn, J., \& Palmer, W.P., 1993, ApJ, 415, L137

\bibitem 1 L\'opez, J. A., Vazquez, R., \& Rodriguez, L.F., 1995, ApJ, 455, 63
 
\bibitem 1 Lovelace, R. V. E., Romanova, M. M., \& Bisnovatyi-Kogan, G. S. 1995, 
MNRAS, 275, 244

\bibitem 1 Luo, D., \& McCray, R., 1991, ApJ, 379, 659

\bibitem 1 Luo, D., Mccray, R., \& Slavin, J., 1994, ApJ, 430, 264

\bibitem 1 Manchado, A, Guerrero, M., Stanghelli, L., \& Serra-Ricart, M., 1996
"The IAC Catalog of Northern Galactic PNe", (IAC, La Laguna)

\bibitem 1 Marston, A. P., Yocum, D. R., Garcia-segura, G., Chu, Y-H.  
1994, ApJS, 95, 151

\bibitem 1 Marten, H. \& Schoenberner, D,  1991, A\&A, 248, 590

\bibitem 1 Martin, C. L., \& Arnett, D, 1995, ApJ, 447, 378

\bibitem 1 Masson, C.~R. \& Chernin, L.~M. 1993, ApJ, 414 230

\bibitem 1 Masson, C.~R. \& Chernin, L.~M. 1992, ApJ, 387, L47

\bibitem 1 McCray, R., Lin, D. N. C 1994, Nature, 369, 378

\bibitem 1 McCaughrean, Mark J., O'Dell, C. Robert, 1996, AJ, 111, 1977

\bibitem 1 Meixner, M., Skinner, C. J., Graham, J. R., Keto, E.,
 Jernigan, J. G., Arens, J. F. 1997, ApJ, 482, 897

\bibitem 1 Mellema, G., Eulderink, F., \& Icke, V., 1991, A\&A, 252, 718.

\bibitem 1 Mellema, G., 1994, A\&A, 290, 915

\bibitem 1 Mellema, G., \& Frank, A., 1995, MNRAS, 273, 40

\bibitem 1 Mellema, G., 1995, MNRAS, 277, 173

\bibitem 1 Mellema, G. 1997, A\&A, 321L, 29

\bibitem 1 Mellema, G., \& Frank, A., 1997, MNRAS, 292, 795

\bibitem 1 G. Mellema, 1996,
in: Jets from Stars and Galactic Nuclei, Springer Lecture Notes, W.R. Kundt (ed.). Springer, Berlin. p.
149. (Springer, New York)

\bibitem 1 Mellema, G., Raga, A, Canto, J., Lundqvist, P., Steffan, W.,
\& Noriega-Crespo, A., 1998, A\&A, 331, 335

\bibitem 1 Meyer, F., 1997, MNRAS, 285, 11

\bibitem 1 Morse, J., Davidson, K,, Ebbets, M., Walborn, N, Balick, B., Frank, A.,
\& Bally, J., 1998, AJ, in press

\bibitem 1 Nota, A., Livio, M., Clampin, M., \& Schulte-ladbeck, R.,
1995, ApJ, 448, 788 (NLCS)

\bibitem 1 Norman, M.L., 1993, in ``Astrophysical Jets'', eds. D. Burgarella,
M. Livio, \& C. O'Dea, Cambridge University Press, 210.

\bibitem 1 O'Dell, C. R., \& Handron, K. D., 1996 AJ, 111, 1630

\bibitem 1 O'Dell, C. R., Weiner, L., \& Chu, Y-H., 1990, ApJ, 362, 2260

\bibitem 1 Oyued R., \& Pudritz, R. E. 1997, ApJ, 482, 712

\bibitem 1 Ostriker, E., 1997, ApJ, 486, 291

\bibitem 1 Osterbrock, D., 1989, Astrophysics of Gaseous Nebulae and Active
Galactic Nuclei, (University Science Books, Mill Valley CA)

\bibitem 1 Owocki, S., Cranmer, S., \& Blondin, J., 1994, ApJ, 424, 887

\bibitem 1 Owocki, S., Cranmer, S., \& Gayley, K. G., 1996 ApJ, 472L, 115

\bibitem 1 Palmer, J. W., Lopez, J. A., Meaburn, J., Lloyd, H. M 1996, A\&A, 307, 225

\bibitem 1 Parker, E., 1979, Cosmical Magnetic Fields, (Oxford University
Press, New York)

\bibitem 1 Pascoli, G., 1985, A\&A, 147, 257

\bibitem 1 Pascoli, G., 1997, ApJ, 489, 946

\bibitem 1 Peter W.,  \& Eichler D., 1996, ApJ, 466, 840

\bibitem 1 Pikel'ner S.B., 1968, Astrophys. ~Lett., 2, 97

\bibitem 1 Podsiadlowski, P., Fabian, A. C., Stevens, I. R  1991 Nature, 354, 43

\bibitem 1 Podsiadlowski, P, 1992, PASP, 104, 717

\bibitem 1 Pollocco, D., L., 1998, in Planetary Nebulae IAU Symp 180,
in press

\bibitem 1 Priest, E.R., 1984, Solar Magnetohydrodynamics, (Reidel, Dordrect)

\bibitem 1 Pudritz, R.E. 1991, in ``The Physics of Star Formation and Early 
Stellar Evolution'', eds. C.J. Lada and N.D. Kylafis, NATO ASI Series 
(Kluwer), 365.

\bibitem 1 Raga, A. C., \& Cant\'o, J.1989, ApJ, 344, 404

\bibitem 1 Raga, A.~C., Cant\'o, J., Calvet, N., Rodriguez, L.~F., \& Torrelles, J.~M.  1993, AA, {\bf 276} 539

\bibitem 1 Raga, A.C., \& Cabrit, S., 1993, A\&A, 278, 26

\bibitem 1 Raga, A.~C., Cant\'o, J., Calvet, N., 
Rodriguez, L.~F., \& Torrelles, J.~M.  1993b, AA, {\bf 276} 539

\bibitem 1 Rasio, F., \& Livio, M., 1996, ApJ, 471, 366

\bibitem 1 Ray, T, Mundt, R., Dyson, J., Falle, S., \& Raga, A. 1996, ApJ, 468L, 103

\bibitem 1 Richling, S., \& Yorke, H. W. 1997, A\&A, 327, 317

\bibitem 1 Rozyczka, M., \& Tenorio-Tagle, G., 1985, A\&A, 147, 209 

\bibitem 1 Rozyczka, M., \& Franco, J., 1996, ApJ, 469, 127

\bibitem 1 Reid M. J., Argon A. L., Masson C. R., Menten K. M.  Moran J. M.,
1995, ApJ, 443, 238

\bibitem 1 Reipurth, B., 1997, in Herbig-Haro Flows and the Birth of
Low Mass Stars, in IAU Symposium no. 182, eds B. Reipurth \& C Bertout,
(Kluwer, Dortdrecht)

\bibitem 1 Reipurth, B, Bally, J, Devine, D., 1997, AJ, 114, 2708

\bibitem 1 Romanova, M., Ustyugova, G., Koldoba, A., Chechetkin, V., \& Lovelace, R., 
1997, ApJ, 482,70

\bibitem 1 Sahai, R., \ea THE WFPC-2 Team, 1998a, ApJ, 1998, ApJ, 493, 301

\bibitem 1 Sahai, R., \ea THE WFPC-2 Team, 1998b, ApJ, 1998, ApJ, 492L, 163

\bibitem 1 Sandquist, E., Taam, R., Chen, X., Bodenheimer, P.,  
Burkert, P., 1998, ApJ, in press

\bibitem 1 Snell, R. L., Loren, R. B., \& Plambeck, R. L., 1980, ApJ, 239, 17
 
\bibitem 1 Schonberner, D., 1986, A\&A, 169, 189

\bibitem 1 Schwarz, H. E., Corradi, R. L. M., Melnick, J. 1993, A\&A, 268, 714

\bibitem 1 Schwarz, H. E., Corradi, R. L. M., Melnick, J. 1992, A\&AS, 96, 23

\bibitem 1 Schmidt-Voigt, M., \& Koeppen, J. 1987A\&A, 174, 211

\bibitem 1 Shepherd, D. S., Churchwell, E. 1996, ApJ, 472, 225

\bibitem 1 Shore, S., 1992, Astrophysical Hydrodynamics
(Academic Press, San Diego)

\bibitem 1 Shu, F., 1997, in Herbig-Haro Flows and the Birth of Low
Mass Stars, in IAU Symposium no. 182, eds B. Reipurth \& C Bertout,
(Kluwer, Dortdrecht)

\bibitem 1 Shu, F., 1992, "The Physics of Astrophysics. Vol.2: Gas Dynamics"
(University Science Books, Mill Valley, CA) 

\bibitem 1 Shu, F.~H., Ruden, S.~P., Lada, C.~J., \& Lizano, S. 1991, ApJ, {\bf 370} L31

\bibitem 1 Shu, F., Najita, J., Ostriker, E., Wilkin, F, Ruden, S.,  Lizano, S., 1994, ApJ, 429, 781

\bibitem 1 Shu, F., Najita, J., Ostriker, E., \& Shang S., 1995, ApJ, 495, L155

\bibitem 1 Smith, M.D., Smarr, L., Norman, M,L., \& Wilson, J.R., 1983
\apj, 246, 432

\bibitem 1 Smith, M.D., MNRAS, 1986, 223, 57

\bibitem 1 Smith, N., \& Ghertz R., 1998, preprint

\bibitem 1 Soker, N., Livio, M., 1989, ApJ, 339, 268

\bibitem 1 Soker, N., Livio, M., 1994, ApJ, 421, 219

\bibitem 1 Soker, N, 1994, AJ, 107, 276

\bibitem 1 Soker, N, 1996, ApJ, 460L, 53

\bibitem 1 Soker, N, 1997, ApJS, 112, 487S

\bibitem 1 Soker N., 1997a in the 10th Cambridge Workshop on Cools Stars, Stellar Systems and the Sun, in press

\bibitem 1 Soker N., 1998, ApJ, in press

\bibitem 1 Stahler, S. 1993, in {Astrophysical Jets, eds. M. Livio, C.~O.~Dea and D. Burgarella, (Cambridge U. Press, New York) 183}

\bibitem 1 Suttner, G. Smith, M. D. Yorke, H. W.\& Zinnecker, H., 1997, A\&A, 318, 595

\bibitem 1 Stanghellini, L., Corradi, R. L. M., Schwarz, H. E., 1993, A\&A, 276, 463

\bibitem 1 Tenorio-Tagle G., Cant{\'o} J., R{\'o}\.zyczka M., 1988, A\&A, 202, 256 (TTCR)

\bibitem 1 Tereby, S., Shu, F. \& Cassen, P., 1984, \apj, 286,529

\bibitem 1 Terman J. L.,. Taam, R.E.,  \& Hernquist, L., 1994, ApJ, 422, 729

\bibitem 1 Terman J. L.,. Taam, R.E.,  \& Hernquist, L., 1995, ApJ, 445, 367

\bibitem 1 Trammell, S., \& Goodrich, R., 1996, ApJ, 468, 107

\bibitem 1 Vishniac, E. T., 1983, ApJ, 274, 152

\bibitem 1 Vishniac, E. T., 1994, ApJ, 428, 186

\bibitem 1 Wang, L., \& Mazzali, P., 1992, Nature, 355, 58

\bibitem 1 Washimi, H, Shibata, S., Mori, M., 1996, PASJ, 48, 23

\bibitem 1 Weaver, R., Mccray, R., Castor, J., Shapiro, P., Moore, R., 1977, ApJ, 218, 377

\bibitem 1 Wilkin, F., \& Stahler, S., 1998, ApJ, in press

\bibitem 1 Weiss, N. O. 1994. "Solar and Stellar Dynamos".In "Lectures on Solar and
Planetary Dynamos", ed. M.R.E. Proctor and A.D. Gilbert. Cambridge:
Cambridge University Press, 59-95

\bibitem 1 Weis, K., Duschl, W., Bomans, D., Chu, Y.-H., \& Joner, M., 1996, 
A\&A, in press

\bibitem 1 Wilner D. J., Reid M. J., Menten K. M., Moran J. M., 1997, 
in Malbet F., Castets A., eds., Poster proceedings of IAU Symp.~182 on
Herbig--Haro Objects and the Birth of Low Mass Stars. Obs.~de Grenoble,
Grenoble. p.~193

\bibitem 1 Yorke, H., \& Welz, A., 1996, A\&A, 315, 555

\bibitem 1 Zimmer, F., Lesch, H., Birk, G.,  1997, A\&A, 320, 746

xxx

\end{thebibliography}
\end{document}